\documentclass[preprint]{aastex6}
\usepackage{xspace,graphicx,verbatim}

\newcommand{\co}[2]{$^{#1}$CO$\;${#2}\xspace}
\newcommand{\CO}[1]{$^{#1}$CO\xspace}

\begin{document}

\title{Structural and Dynamical Analysis of 0.1pc Cores and Filaments in the 30~Doradus-10 Giant Molecular Cloud}
\author{
R\'emy Indebetouw\altaffilmark{1,2},
%Crystal Brogan\altaffilmark{1},
T. Wong\altaffilmark{3},
C.-H. Rosie Chen\altaffilmark{3},
%Adam Leroy\altaffilmark{1},
%Kelsey Johnson\altaffilmark{2},
Amanda Kepley\altaffilmark{1},
Vianney Lebouteiller\altaffilmark{4},
Suzanne Madden\altaffilmark{5},
%Diane Cormier\altaffilmark{7},
%Annie Hughes\altaffilmark{8},
%Todd Hunter\altaffilmark{1},
%Akiko Kawamura\altaffilmark{6},
%Margaret Meixner\altaffilmark{9},
Joana M. Oliveira\altaffilmark{6},
%Toshikazu Onishi\altaffilmark{11},
}
\altaffiltext{1}{National Radio Astronomy Observatory, 520 Edgemont Road
Charlottesville, VA 22903; rindebet@nrao.edu, akepley@nrao.edu} 
\altaffiltext{2}{Department of Astronomy, University of Virginia, P.O. Box 3818, Charlottesville, VA 22903-0818; remy@virginia.edu}
\altaffiltext{3}{Department of Astronomy, University of Illinois, 1002 W Green St, Urbana, IL 61801; wongt@illinois.edu}
%\altaffiltext{3}{Max-Planck-Institut f\"ur Radioastronomie, Auf dem H\"ugel 69, D-53121, Bonn, Germany; rchen@mpifr-bonn.mpg.de}
%\altaffiltext{6}{ALMA-J Project Office, National Astronomical Observatory of Japan, 2-21-1 Osawa, Mitaka, Tokyo 181-8588, Japan; erik.muller@nao.ac.jp, akiko.kawam
%ura@nao.ac.jp}
\altaffiltext{4}{AIM, CEA, CNRS, Universit\'e Paris-Saclay, Universit\'e Paris Diderot, Sorbonne Paris Cit\'e, F-91191 Gif-sur-Yvette, France; vianney.lebouteiller@cea.fr}
\altaffiltext{5}{IRFU, CEA, Universit\'e Paris-Saclay, F-91191 Gif-sur-Yvette, France; suzanne.madden@cea.fr}
%\altaffiltext{7}{Institut f\"ur theoretische Astrophysik, 
%Zentrum f\"ur Astronomie der Universit\"at Heidelberg, 
%Albert-Ueberle Str. 2, D-69120 Heidelberg, Germany; diane.cormier@zah.uni-heidelberg.de}
%\altaffiltext{8}{Max-Planck-Institut f\"ur Astronomie, K\"onigstuhl 17, D-69117, Heidelberg, Germany; hughes@mpia.de}
%\altaffiltext{9}{Space Telescope Science Institute, 3700 San Martin Drive, Baltimore, MD 21218; meixner@stsci.edu}
\altaffiltext{6}{School of Physical and Geographical Sciences, Lennard-Jones Laboratories, Keele University, Staffordshire ST5 5BG, UK; j.oliveira@keele.ac.uk}
%\altaffiltext{11}{Department of Physical Science, Graduate School of Science, Osaka Prefecture University, 1-1 Gakuen-cho, Naka-ku, Sakai, Osaka 599-8531, Japan; ohnishi@p.s.osakafu-u.ac.jp}

\begin{abstract}
  High-resolution ($<$0.1\,pc) ALMA observations of the 30Dor-10 molecular
cloud 15\,pc north of R136 are presented.  The \co{12}{2-1} emission
morphology contains clumps near the locations of known mid-infrared
massive protostars, as well as a series of parsec-long filaments
oriented almost directly towards R136.
There is elevated kinetic energy (linewidths at a given size scale) in 30Dor-10 compared to other LMC and Galactic star formation regions, consistent with large scale energy injection to the region.
Analysis of the cloud substructures is performed by segmenting emission into disjoint approximately round ``cores'' using {\tt clumpfind}, by considering the hierarchical structures defined by isointensity contours using {\tt dendrograms}, and by segmenting into disjoint long thin ``filaments'' using {\tt Filfinder}. Identified filaments have widths $\sim$0.1\,pc.
The inferred balance between gravity and kinematic motions depends on the segmentation method: 
Entire objects identified with {\tt clumpfind} 
are consistent with free-fall collapse or virial equilibrium with moderate external pressure, whereas many dendrogram-identified parts of hierarchical structures have higher mass surface densities $\Sigma_{LTE}$ than if gravitational and kinetic energies were in balance.
Filaments have line masses that vary widely compared to the critical line mass calculated assuming thermal and nonthermal support.  Velocity gradients in the region do not show any strong evidence for accretion of mass along filaments.
The upper end of the ``core'' mass distribution is consistent with a power-law with the same slope as the stellar initial mass function.

\end{abstract}

\section{Introduction}
\label{introduction}

One major open question in star formation is how gas is accreted from
the clump and cloud scale ($\sim$10\,pc) onto nascent stars.
For many years there has been debate about whether isolated cores
($\sim$0.1\,pc) form which then collapse into individual stars and
multiples, or whether a significant fraction of the eventual stellar
mass is accreted from further away in the cluster gravitational
potential well (often referred to as competitive or collaborate
accretion). A related question is whether the stellar initial mass
function is already set by cloud fragmentation into the core mass
function, before most of the matter has accreted onto protostars
\citep[e.g.][and refs therein]{motte_araa}.
Over the last decade, it has become clear that in solar neighborhood clouds,
filamentary structures ($\sim$0.1\,pc$\times$1\,pc) are ubiquitous, contain most of the dense cores in such
clouds, and likely play a role in transferring matter from
the cloud to core scales \citep[e.g.][]{andre_fils,classy_fils}.

To develop any universal understanding of star formation it is
important to determine whether the same core and filament structures
exist in a wider range of molecular clouds than exist in the solar
neighborhood.  Distant regions in the Milky Way disk suffer from both
significant distance (and thus source luminosity) uncertainty, and
confusion along the line of sight.  The Large Magellanic Cloud (LMC)
is not much farther, but being close to face-on suffers from little
distance uncertainty or line of sight confusion.  The 30Dor-10
molecular cloud \citep{johansson98} is located 15pc to the NE of the
rich star cluster R136 in the LMC.  The ultraviolet (UV) radiation
field affecting 30Dor-10 is 3500$\times$ that of the solar
neighborhood \citep{werner78}, and the reduced dust abundance at the
region's $\sim$1/2 solar metallicity \citep{russell92,peimbert03}
makes the interstellar medium more permeable to that radiation
\citep{poglitsch}.  This makes 30Dor-10 a good laboratory in which to
study star formation in the presence of strong external radiative
feedback.  30$\;$Doradus is also remarkable compared to typical star
formation regions in the Magellanic clouds or Milky Way, and perhaps
signatures of what caused such an intense star formation event are
still evident in the structure of the residual molecular gas.  Low and
intermediate-mass star formation is actively ongoing in 30Dor-10
\citep{rubio98,walborn13,sabbi16} and now with the Atacama Large
(sub)Millimeter Array (ALMA) we can obtain a detailed picture of the molecular gas down to 0.1pc scales.

In ALMA Cycle 0 we mapped the 30Dor-10 cloud in \co{12}{2-1}, \co{13}{2-1},
C$^{18}$O\,2-1, and 1.3\,mm continuum
at a resolution of
$\simeq$2.3\,\arcsec$\times$1.5\,\arcsec \citep[2011.0.00471.S,][hereafter Paper I]{ri13}, equaling = 0.56\,pc$\times$0.36\,pc at the adopted distance of 50\,kpc \citep[e.g.][]{distance}.  We found that resolved parsec-scale structures have $>$3$\times$ larger velocity linewidths at those scales than other star formation regions in the Milky Way and LMC \citep{ri13,nayak16}. This can be explained by an external pressure of P$_e$/k$>$10$^6$\,cm$^{-3}$K, either from the bubble and ionized regions around R136, or merely from the weight of the molecular cloud envelope in which the observed clumps are embedded.  The slope of the size-linewidth relation agreed with other regions within uncertainties, and there were no strong trends of clump property with distance from R136, so the question of whether star-forming clumps in 30Dor-10 are affected by feedback, or that the entire region is simply very turbulent, was unresolved.  This study extends the previous work to much smaller angular resolution, now resolving the 0.1\,pc-sized cores that are expected to be actively participating in individual or multiple star formation.  The greater spatial dynamic range permits more robust determination of size-linewidth-mass relations, and a core mass function more directly relevant to star formation can be measured, as described in the following sections.

\section{Observations and Data Reduction}
\label{observation}

The 30Dor-10 cloud was observed with ALMA in \co{12}{2-1}, \co{13}{2-1},
C$^{18}$O\,2-1, 1.3\,mm continuum, the H30$\alpha$ recombination
line, H$_2$CO$\;3_{0,3}-2_{0,2}$, $3_{2,2}-2_{2,1}$ and
$3_{2,1}-2_{2,0}$ (218.22219, 218.47563, 218.76007GHz; formaldehyde data will be presented separately).
% 218.22219 is 3(0,3)-2(0,2) E_U=20.956K, other two have E_U=68K
% spws 27, 25, 21, 29, 19, 17, 23
% 218.222192 3(0,3) - 2(0,2) A=-3.55037  10.48357 	20.95665
% 218.475632 3(2,2) - 2(2,1) A=-3.09140  57.60929 	68.09454
% 218.760066 3(2,1) - 2(2,0) A=-3.09030  57.61274 	68.11164
Most of the cloud was mapped as part of project 2011.0.00471.S
\citep{ri13} at a resolution of
$\simeq$2.3\,\arcsec$\times$1.5\,\arcsec = 0.56\,pc$\times$0.36\,pc.
The
brightest parts of the cloud were observed in project
2013.1.00346, targeting the same lines at higher angular ($<$0.1\,pc) and spectral
resolution. (C$^{18}$O$\;$2-1 and the H$_2$CO lines were observed at
122\,kHz=168\,m/s resolution, \co{12}{2-1} and \co{13}{2-1} at
61\,kHz$\simeq$80\,m/s resolution.)
The observation was executed 7 times between 2015/06/27 and
2015/09/24.  The phase calibrator was J0635-7516 (0.46-0.53\,Jy at
230.5\,GHz during the time range of observations).  J0635-7516 was
also used as bandpass calibrator in all but one execution which used
J0538-4405 (1.2\,Jy at 230.5\,GHz).  The amplitude calibrator was
J0519-454 (0.66-1.0\,Jy at 230.5\,GHz).
%bp  J0635-7516 	06:35:46.508 	-075.16.16.815 0.53,0.50,0.46,0.48,0.50
%    J0538-4405 	05:38:50.362 	-044.05.08.939 1.16, 
%amp J0519-454 	05:19:49.723 	-045.46.43.853 1.0, 0.89,0.66,0.66,0.69
%ph  J0635-7516
Data were calibrated using the ALMA Calibration pipeline version
Cycle3R1 included in Common Astronomy Software Applications (CASA;
\url{http://casa.nrao.edu}) v. 4.3.1 \citep{casa}.  Calibrated visibilities were
subsequently continuum-subtracted in the uv domain and deconvolved
using {\tt clean}, {\tt tclean} in CASA 4.5, and a prototype version
of the auto-multithresh automasking routine now included in CASA
\citep[since v. 5.1.0][]{kepley20}.
Visibilities from both projects were included in
the deconvolution.  
The \co{12}{2-1} and \co{13}{2-1} data were imaged at
0.4\,km\,s$^{-1}$ resolution with Briggs weighting, robust=0.5,
and multi-scale deconvolution at 0,5, and 9 times the
0.032$\;$arcsec pixel, achieving a beam of
%0.375$\times$0.25$\;\arcsec$ = 0.09$\times$0.06$\;$pc, and rms of
%1.6$\;$mJy$\;$bm$^{-1}$ and 4.5$\;$mJy$\;$bm$^{-1}$, respectively.
%chev
0.31\,$\arcsec\times$0.22\,$\arcsec$ = 0.08\,pc$\times$0.05\,pc, and rms noise in line-free channels of 2.3mJy$\;$bm$^{-1}$.  The \co{12}{2-1} image was feathered with APEX single dish data to recover all of the $\sim$25\% of the large-scale emission that was resolved out by the interferometer; the combined image used for subsequent analysis has an rms of 6mJy\,bm$^{-1}$.  Comparison of the ALMA and APEX data for \co{13}{2-1} indicates that the interferometer recovered all emission to within uncertainties, so the ALMA-only image is used for \co{13}{2-1} analysis.
The C$^{18}$O$\;$2-1 data were imaged at 0.4\,km\,s$^{-1}$ resolution with
natural weighting, achieving a beam of 0.44\,$\arcsec\times$0.28\,$\arcsec$ and
rms = 1.6\,mJy\,bm$^{-1}$

%XXX H$_2$CO?  various tapers

%=====================================================================
\section{CO morphology and associated star formation}
\label{individuals}

\begin{figure}
  \includegraphics[width=0.7\textwidth]{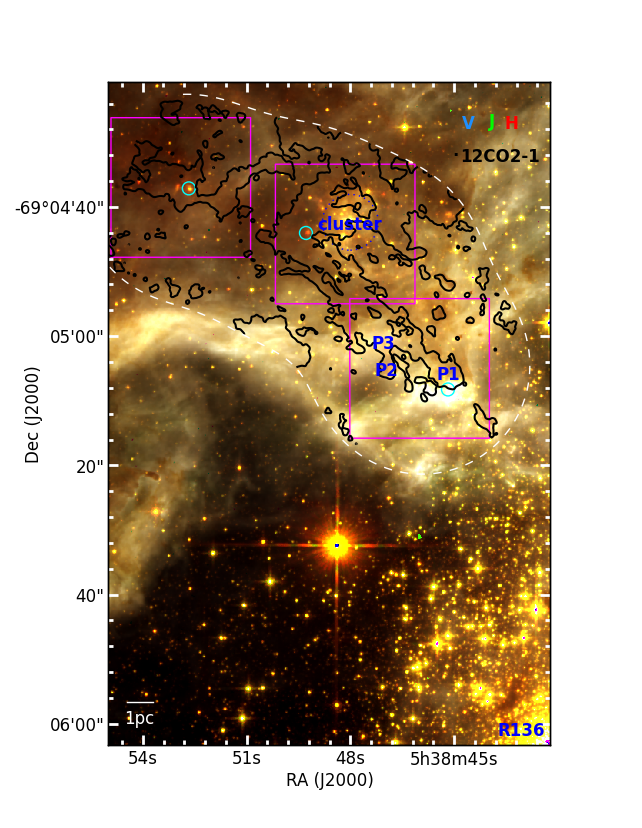}
  \caption{The black contour shows the 3$\sigma$ \co{12}{2-1} peak intensity 30Dor-10, overlaid on a RGB composite of HST F160W$\sim$H, F110W$\sim$J, and F658N (H$\alpha$).  The CO beam is the small black dot to the left of the ``12CO2-1'' label. The dominant sources of ionizing radiation are the stars to the southwest including R136.
    %and the photoionized cloud edge is brightly outlined in H$\alpha$.
    Pillar structures P1,2,3, and a small cluster in the center of the cloud, are labeled in blue, and the three Spitzer-identified massive YSOs in this region are marked in cyan -- see text for discussion.  Figure~\ref{CO_12_13_fig} shows the CO structure in more detail.
    Figure~\ref{mysos} shows zooms of the three regions in magenta.
    \label{CO_HST_fig}}
\end{figure}

\begin{figure}
  \includegraphics[width=0.7\textwidth]{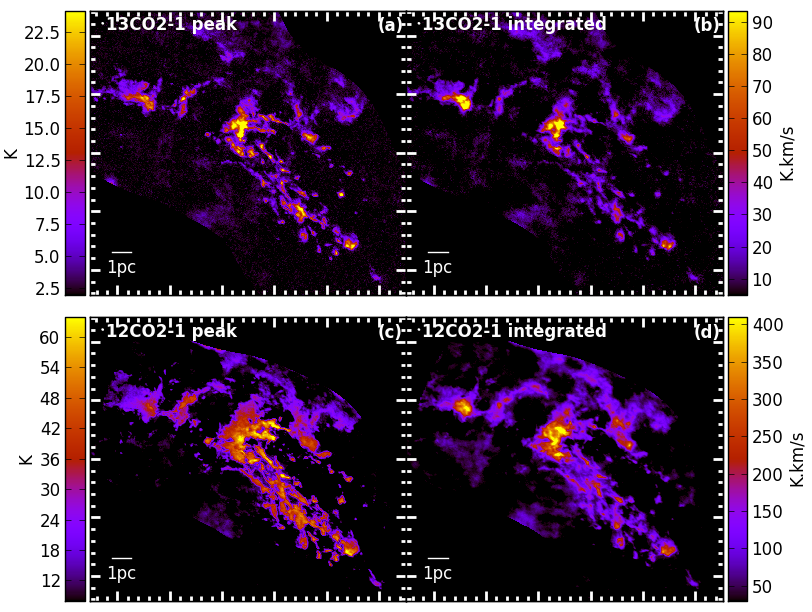}
  \caption{Peak and integrated \co{13}{2-1} and \co{12}{2-1} at 0.1-pc resolution in 30Dor-10 (beams are small white dots to the left of each label).\label{CO_12_13_fig}}
\end{figure}

\begin{figure}
{\includegraphics[width=6cm,trim={6ex 6ex 6ex 6ex},clip]{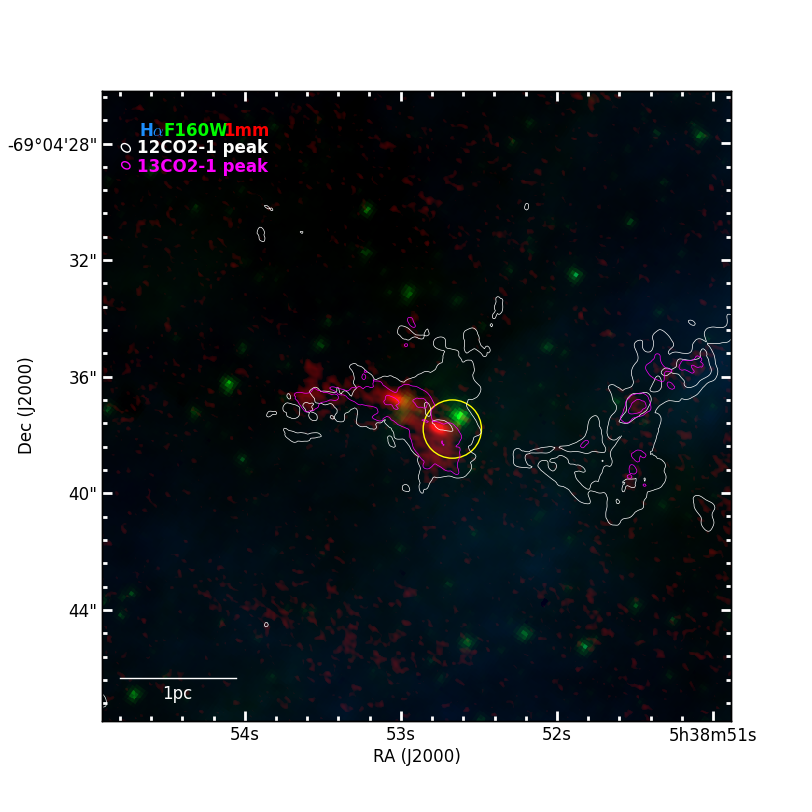}}
{\includegraphics[width=6cm,trim={6ex 6ex 6ex 6ex},clip]{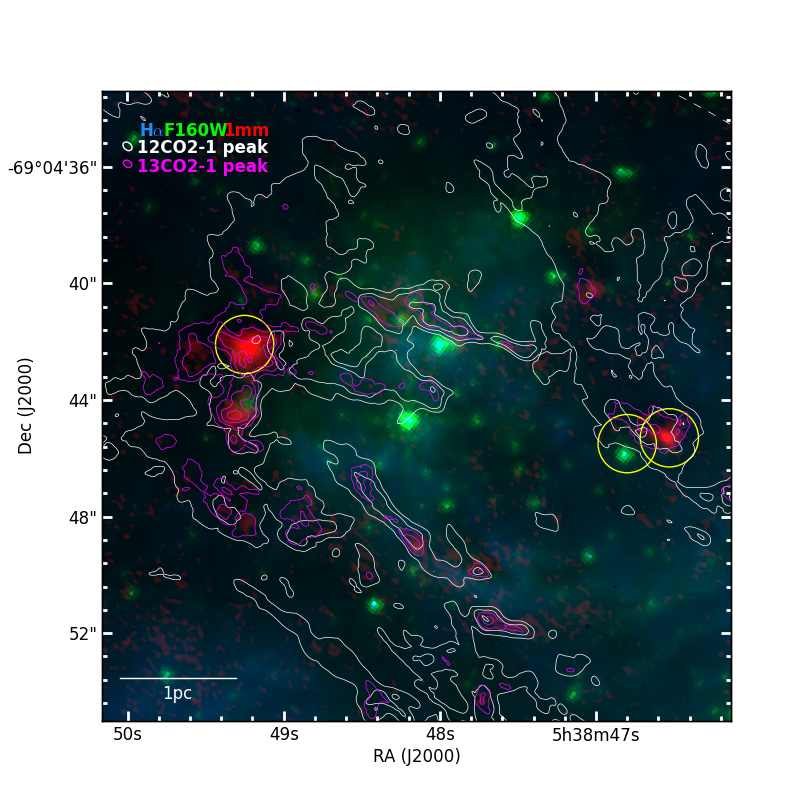}}
{\includegraphics[width=6cm,trim={6ex 6ex 6ex 6ex},clip]{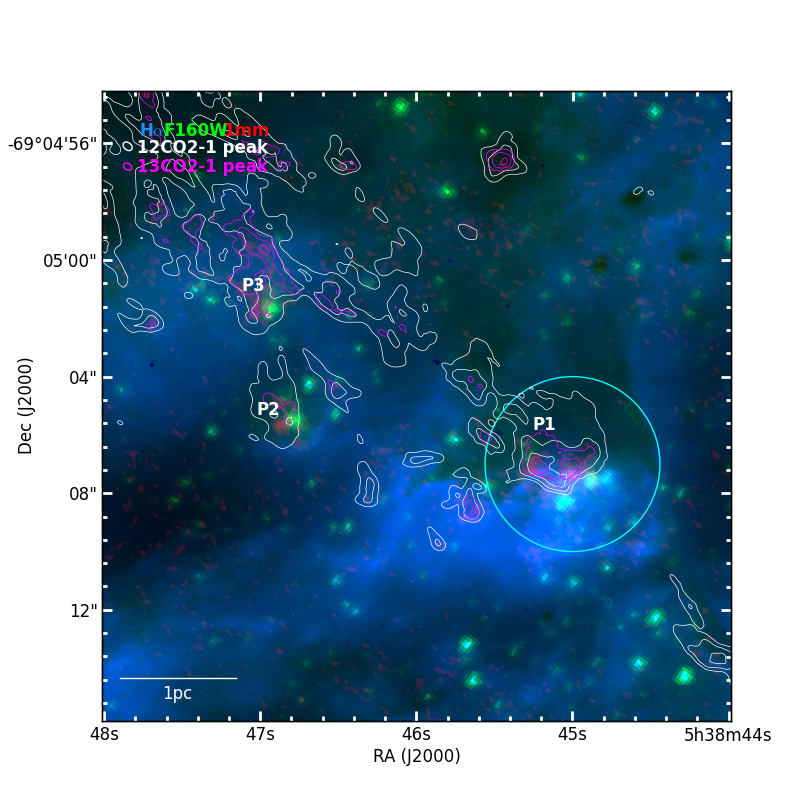}}
  \caption{Zooms on three sections of 30Dor-10, with the same image stretch and contour levels.  Colors are red: 1mm ALMA continuum in square root scale, green: F160W HTTP HST in log scale, blue: F658N (H$\alpha$) HTTP HST in log scale, white contours: \co{12}{2-1} peak brightness at 0.125,0.2,0.275 Jy$\;$bm$^{-1}$ (cube rms = 3.6mJy$\;$bm$^{-1}$), magenta contours: \co{13}{2-1} beak brightness at 0.04,0.06,0.08,.1 Jy$\;$bm$^{-1}$ (cube rms = 2.2mJy$\;$bm$^{-1}$).  H$_2$O masers are marked with yellow circles of radius=1\arcsec, the quoted positional uncertainty \citep{imai}.  An OH maser is marked similarly with a cyan circle \citep{brogan_maser}.}\label{mysos}
\end{figure}

Figure~\ref{CO_HST_fig} shows the extent of \co{12}{2-1} emission relative to optical and near-infrared emission imaged with HST for the HTTP project \citep{sabbi}, and the location of features that will be discussed in this section.
Figure~\ref{CO_12_13_fig} shows the peak and integrated emission of \CO{12}{2-1} and \CO{13}{2-1}, and Figure~\ref{mysos} shows three zoomed-in subregions in F160W$\sim$H, F110W$\sim$J, and 1mm continuum.

Nearest R136 in the southwest, the CO-traced molecular gas is characterized by photodissociated pillars (P1,2,3 in Figure~\ref{CO_HST_fig}) each of which has embedded 1mm sources visible in Figure~\ref{mysos}, third panel.
Interestingly, the pillar ``P1'' closest to R136 in projection happens to be the largest sized pillar, and contains two very embedded YSOs, only detected in the millimeter continuum.  Just in ``front'' (SW, i.e. on the R136-side) of the pillar are two near-infrared-detected, more exposed or evolved, stars.  This sequence of more embedded sources being further back from the photodissociating source has been studied in the Milky Way, and cited as evidence for triggered star formation by compression of the pillar heads, but causality is challenging to establish \citep{fukuda_M16, dale_trigger}.
A 1720GHz OH maser was detected at the pillar position \citep[Figure~\ref{mysos}, third panel;][]{brogan_maser}, but those data had insufficient angular resolution to associate the maser with the embedded protostars, or the outer part of the pillar compressed by the HII region.  1720GHz OH masers are often associated with supernova remnant-molecular cloud interactions, and a similar shock could be being driven into the pillar, but such masers are also associated with massive YSOs \citep[e.g.][]{gray92}.
Pillar 1 is also associated with one of three ``massive YSOs'' identified in this region with Spitzer \citep{whitney,grundl,walborn13}. However, even with IRAC at $\leq$8$\mu$m, Spitzer's resolution is nearly a half-parsec, so its not possible to distinguish a small cluster from an individual massive protostar.  With HST's resolution of $\sim$15000AU, resolved red near-infrared sources at the positions of the red Spitzer sources can be much more convincingly called single or multiple MYSOs, and now with ALMA the small-scale structure of their associated molecular gas is resolved. The Pillar 1 ``massive YSO'' is clearly a small group including the more embedded millimeter continuum sources and the more exposed NIR sources in the bright photoionized rim.

Slightly further away from R136 in projection ($\sim$3pc) are two other pillars ``P2,3'' which each contain a YSO visible in the NIR.  Both embedded sources are slightly extended in NIR and 1mm continuum.

The main part of the region, between the south-west pillars and the central cluster and clumps, is characterized by multiple narrow filaments pointing approximately at the central stellar cluster R136.  Photoionization can erode clouds leaving shadowed structures and pillars pointing towards the radiation source \citep{grit10}, but it is also possible that this structure is instead related to the formation of 30~Doradus: \citet{rahner18} model the region as an older burst of star formation, followed by recollapse and formation of R136, at high molecular cloud mass and density.  Although their model is only 1-dimensional, repeated feedback-driven recollapse and non-spherical re-expansion could naturally lead to dense radial structures.

In the center of the mapped region is a small ($\sim$1.5pc diameter, see Figure~\ref{CO_HST_fig}) intermediate-mass cluster associated with diffuse H$\alpha$ emission, and at least one very massive star \citep[O3-6;][]{wb97}.
%which has Wolf-Rayet lines detected in a large optical spectroscopic aperture.  This suggests that the cluster contains at least one massive star and is at least 3Myr old.
Remnant filamentary CO-traced molecular gas is coincident with the eastern side of the cluster, likely in front of the cluster, since the CO filaments correspond to shadows in the diffuse H$\alpha$. These filaments may be under compression by the expanding ionized region, but there is no evidence in millimeter continuum or NIR of embedded protostars.  Just east of the cluster is the brightest CO clump in our data, now resolved into a southern core with an embedded near-infrared source, and a northern ``hub'' with filaments leading into it (although there are no clear kinematic signatures of accretion along those spokes).  The northern source has an extended 1mm source, not detected in the NIR, and is likely the young MYSO that dominates the MIR emission from the region (Spitzer cannot resolve these two MYSOs).  That more embedded northern source is associated with a H$_2$O maser \citep{imai}, thought to be collisionally pumped, and most commonly associated with massive YSO outflows \citep{elitzur,walsh}.
The ``hub-and-spoke'' morphology of the CO emission is also commonly seen in Galactic MYSOs forming in molecular clouds.  Thus, although the source is suggestively located on the edge of the small HII region, it is difficult to unambiguously claim that its formation was triggered by that HII region.
On the western side of the small cluster is another CO clump coincident with two detected H$_2$O masers.  One is at the position of a 1mm continuum source, the other might coincide with a NIR-detected star.

On the eastern side of our mapped area, furthest from R136, is another very massive clump, with at least one deeply embedded massive protostar detected in 1mm continuum.  The kinematic structure is complex, and suggestive of multiple outflows.  It is not clear whether the near-IR source is an illuminated outflow cavity from the massive protostar(s) or an unassociated source.

% VELOCITY OF FILS?  IN PILLARS?  at least is sigma high?

%=====================================================================

\section{CO$\;$2-1 Structure Analysis}
\label{analysis}

\subsection{Methods}

Molecular clouds have complex hierarchical structure on a large range of spatial scales, but there is insight to be gained by segmenting the emission and analyzing it as discrete structures.  The most common technique has been segmentation of position-position-velocity cubes into approximately round entities a few times the spatial resolution - these are typically called clumps or cores, for $\sim$parsec-sized, and $\sim$0.1\,pc-sized entities, respectively.  The segmentation process begins by finding local maxima, with two critical parameters: $\delta I$ the intensity difference between a local maximum and highest connected saddle point, and $I_{min}$ the minimum intensity to consider.  Neighboring pixels are assigned to each local maxima to define a set of clumps.  Popular implementations are {\tt clumpfind} \citep{clumpfind,clumpfind_ascl}, which assigns all emission down to $I_{min}$, and {\tt cprops} \citep{cprops,cprops_ascl}, which only assigns emission down to the lowest isointensity surface that contains a single local maximum.
%Alternate clump segmentation methods include fitting Gaussians \citep[{\tt gaussclumps}][]{gaussclumps,gaussclumps_ascl}, and other watershed and gradient methods \citep{cupid_ascl}.

An alternate analysis considers isointensity surfaces as a set of hierarchical entities, instead of assigning emission to disjoint clumps.  This is more naturally suited to the hierarchical structure in molecular clouds, but a given emitting pixel gets plotted and analyzed multiply, as part of multiple isointensity structures. The smallest structures associated with local intensity maxima are called ``leaves'', and the largest isolated regions of emission called ``trunks'' or ``islands''. The most commonly used implementation is {\tt dendrograms} \citep{dendro}.
%Various methods have been proposed to cluster the hierarchical entities into disjoint ones based on measured properties such as silhouette or velocity extent \citep[e.g.][]{scimes}.
%
For clump segmentation, we use the {\tt quickclump} python implementation (\url{https://github.com/vojtech-sidorin/quickclump/}) to which we added a $I_{minpk}$ parameter, the minimum peak intensity required for a valid clump (\url{https://github.com/indebetouw/quickclump}).  This allows the fainter envelopes of bright clumps to be included without keeping faint noise peaks.  Properties (moments, fitted sizes, etc) of clumps are calculated with a python translation of the moments calculations in {\tt cpropstoo} (\url{https://github.com/akleroy/cpropstoo}).  For dendrograms we use the python implementation at \url{https://github.com/dendrograms/astrodendro}.

\begin{figure}
  \includegraphics[width=0.44\textwidth]{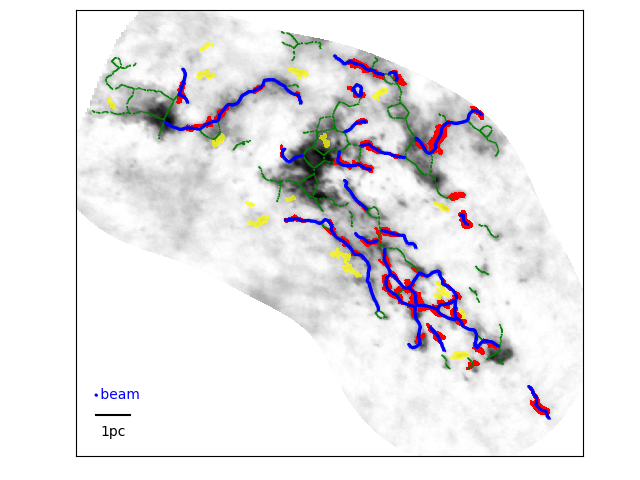}
  \includegraphics[width=0.44\textwidth]{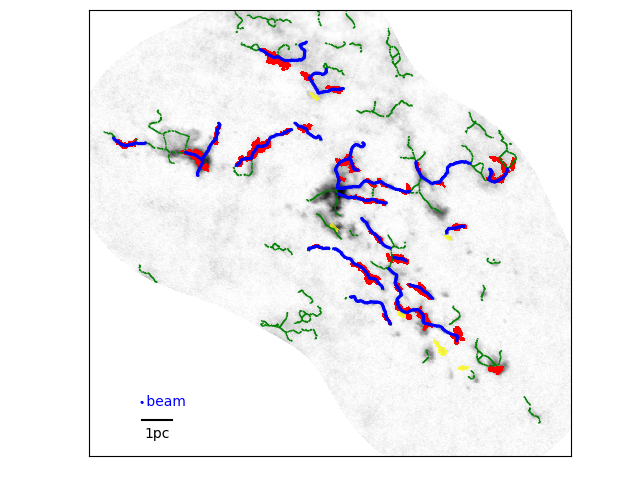}
  \caption{Filaments identified in peak intensity maps of \co{12}{2-1}(L) and \co{13}{2-1}(R).  Filaments that match with elongated dendrogram structures are shown in blue, with their corresponding dendrogram structures in red.  Filaments that don't match elongated dendrogram structures are marked in green. Elongated dendrogram structures that don't match filaments are shown in yellow.
        \label{fils_dendro_fig}}
\end{figure}

Somewhat more recently it has become popular to identify elongated regions of emission and refer to them as filaments.
%Several mathematical methods have been demonstrated including wavelets \citep[{\tt getfilaments}][]{getfilaments}, critical manifolds \citep[{\tt DISPERSE}][]{disperse}, or Hessian operators \citep{salji15}.
%We experimented with {\tt DISPERSE} but settled on {\tt filfinder} \citep{filfinder}, which uses a Medial Axis Transform on the masked image. We could get comparable results with {\tt DISPERSE} and {\tt filfinder}, but found the latter easer to use,
We use {\tt filfinder} \citep{filfinder} and its python implementation (\url{https://github.com/indebetouw/FilFinder}) which allowed for easy modification, testing, and incorporation into other analysis scripts.  Filaments are initially identified in the (2D) peak intensity image - if there is not an overabundance of sightlines with multiple velocity components, this should work well, and indeed for this data, the method produces a visually very satisfactory result (Figure~\ref{fils_dendro_fig}).
Manual examination and searching of this data cube only finds one position at which there are two bright structures at different velocities, and fewer than 10 positions with a bright structure at one velocity, and structure wings or faint emission at another velocity.
{\tt filfinder} performs pruning of small and multiply-connected filament branches.  
We modified {\tt filfinder} to do that pruning in 3D as well as the default 2D.  That requires choosing a metric to calculate length in position-position-velocity space, so we used 1km$\;$s$^{-1}\sim$0.036pc.  Our trials found that the final filament skeletons differ only in minor ways between the default 3D pruning and pruning based on 3D lengths, so here we present only the default 2D results using the publicly available code.

% pix 0.0121pc, 0.3333km/s

It is important to note the bias of the various algorithms:  clump segmentation will divide emission into approximately round entities a few times the beam size.  Filament finders will identify filament skeletons for any distribution of emission, whether visually filamentary or not.  Dendrograms do not impose such geometric constraints, but multiply-count most emission.  To leverage the objective nature of dendrograms in filament analysis, we match filament branches to elongated dendrogram entities. We identify and place greater confidence in those filaments for which there is an isointensity surface that surrounds most of the filament, which contains most of only one filament, and is elongated (in the end a cutoff aspect ratio $>$ 2.5:1 was used, but higher cutoffs up to 5:1 yielded similar results).  Figure~\ref{fils_dendro_fig} shows the {\tt filfinder} filaments identified in the peak intensity image, and the elongated isointensity surfaces that match some of the filament branches.  Filaments are found using a global threshold of 0.17 and 0.07 times the peak intensity for \CO{12} and \CO{13}, respectively.  The image is flattened at 90\% of peak, smoothed to 3 times the beamwidth, and size and adaptive thresholds of 24 times the beam area and 14 times the beamwidth used (see {\tt filfinder} documentation for parameter descriptions). Branches were pruned with thresholds of 5 beams and 10 pixels.
Only isointensity structures more elongated than 2.5:1 with areas between 0.2 and 0.5 square parsecs were matched to filament branches.
\label{fils_dendro_section}
Overall, there is good but not perfect agreement between identified filaments and elongate isointensity contours, so a filament analysis will reveal different information relative to analyzing all dendrogram structures (defined by isointensity surfaces).  In regions where filaments are relatively isolated, the same filaments are identified using $^{12}$CO and $^{13}$CO, although sometimes the $^{13}$CO intensity is low enough to cause gaps or breaks in the $^{12}$CO-identified filament.
In dense clumps such as those in the center and east of the region, containing the most massive {\it Spitzer}-identified protostars (see \S~\ref{individuals}), $^{12}$CO becomes very optically thick, and the relatively flat spatial intensity profile over the entire clump causes the filament finder to break the clump up into loops or cells (see especially the central clump with cyan loops in the $^{12}$CO image, first panel of Figure~\ref{fils_dendro_fig}.

\subsection{Mass calculation}

Physical analysis of structures identified in clouds requires calculating the mass of each structure.  We calculate the mass per PPV pixel from \CO{12} and \CO{13} brightness using what is sometimes referred to as the LTE or ``standard'' method \citep{bourke97,ri13}. The \CO{12} excitation temperature is derived from the \co{12}{2-1} brightness temperature.  The \CO{13} excitation temperature is assumed to be the same as \co{12}{2-1}, to calculate the \CO{13} optical depth and column density.

To assess the accuracy of this method, we ran a grid of non-LTE excitation models with Radex \citep{radex}, spanning N(\CO{12})$\in$[10$^{16}$,10$^{21}$]cm$^{-2}$, T$_K\in$[2,100]K, n(H$_2$)$\in$[10$^2$,10$^7$]cm$^{-3}$, and N(\CO{12})=75 N(\CO{13}) \citep{heikkila,nikolic}.  From the Radex-computed brightness temperatures for \co{12}{2-1} and \co{13}{2-1}, we apply the standard calculations:
\[ T_{ex} =
{{11.1K}\over{\ln\left({{11.1}\over{I_{12}+0.19}}+1\right)}},\label{Tex}\] where
I$_{12}$ is the \co{12}{2-1} intensity in K.
\begin{eqnarray*}
\tau_0^{13} &=& -\ln\left[1-{{T_B^{13}}\over{10.6}}\left\{ {1\over{e^{10.6/T_{ex}}-1}} - {1\over{e^{10.6/2.7}-1}} \right\}^{-1} \right] \\
N(^{13}{CO}) &=& 1.5\times 10^{14} { {T_{ex} e^{5.3/T_{ex}}\int \tau^{13}_v dv}\over{1-e^{-10.6/T_{ex}}}} 
\label{NCO}
\end{eqnarray*}

\begin{figure}
  \includegraphics[width=0.5\textwidth]{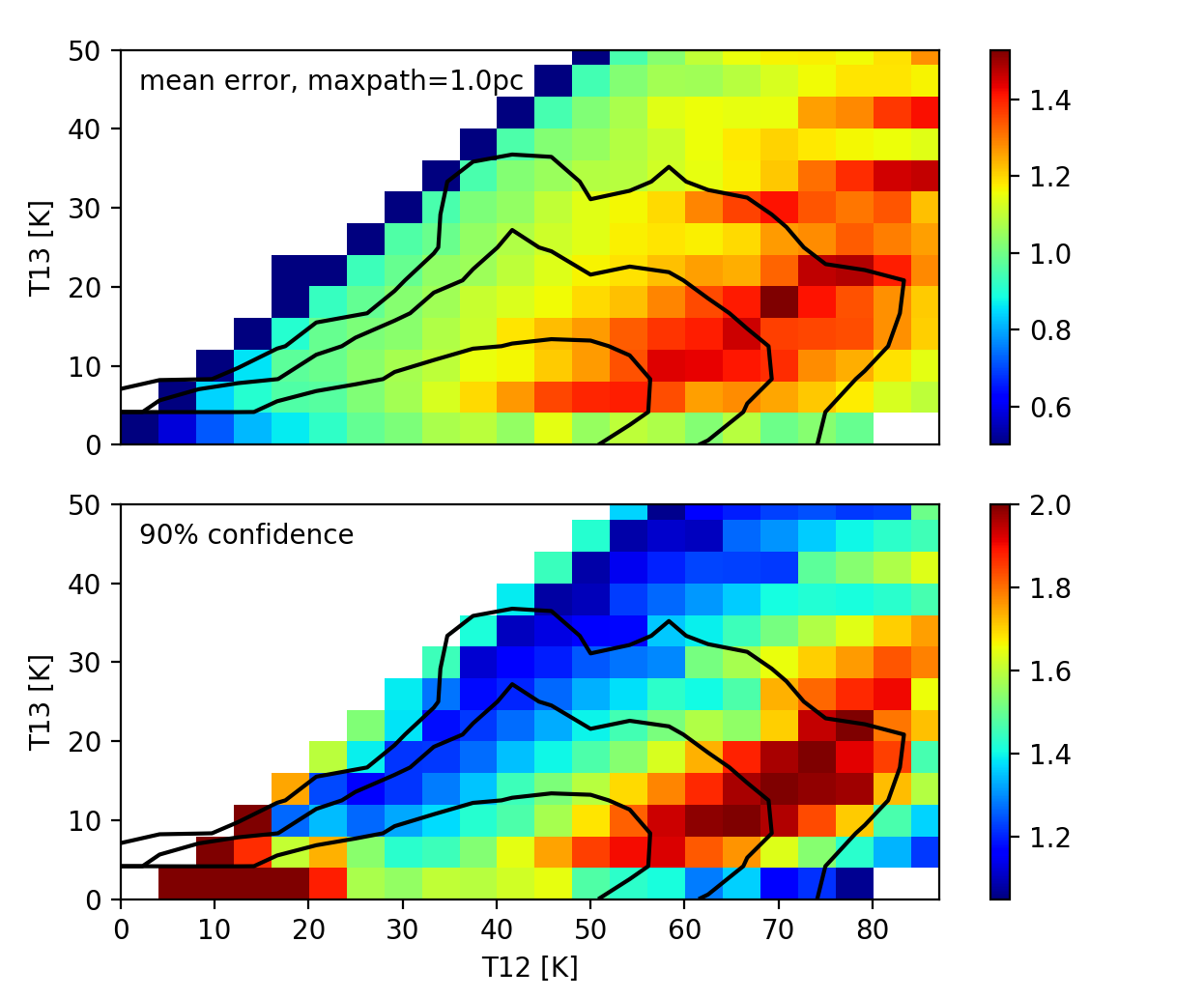}
  \caption{\label{Nrecoveryfig} Assessment of validity of the commonly used ``LTE method'': N(\CO{13}) calculated from modeled $^{12}$CO and $^{13}$CO brightness temperatures, compared to the true values input to each model.
    The top panel is the average N$_{LTE}$/N$_{true}$ for all models with given \co{12}{2-1} and \co{13}{2-1} brightness temperatures.  The contours are a histogram of the observed values in our cubes.  The lower panel shows the value of the error for which 90\% of the models are closer to the true value.  The method works fairly well, overestimating the true column density by up to a factor of 2 at 90\% confidence.}
\end{figure}

We considered only models with N(\CO{13})$<$10$^{14}$n(H$_2$), corresponding to line-of-sight pathlength $<$1pc for an abundance H$_2$/\CO{13}=5$\times$10$^5$.  The results of this section are insensitive to path lengths and H$_2$/\CO{12}/\CO{13} abundances differ by up to a factor of 3 in either direction.  Figure~\ref{Nrecoveryfig} compares N(\CO{13}) calculated with the LTE method to the actual value input to each model. The method works fairly well, with a tendency to overestimate the true column density in brighter regions. There is a modest effect that the calculated column density (and mass) of bright structures may be overestimated by up to a factor of 2 relative to faint structures.
\label{Nrecoverysection}

\subsection{Structure analysis result: filament velocity structure}

\begin{figure}
  \includegraphics[width=0.5\textwidth]{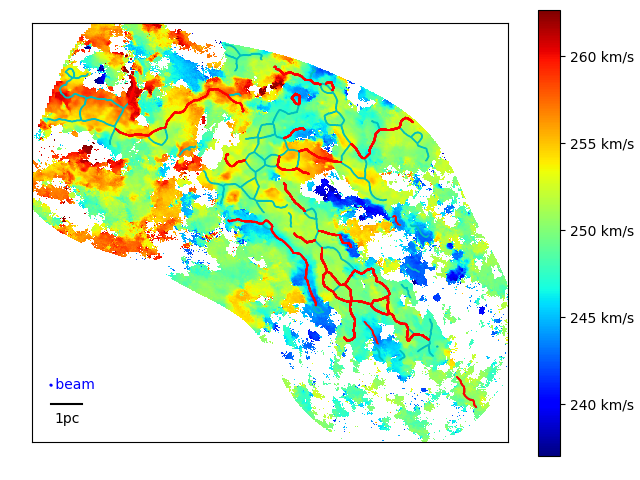}
  \includegraphics[width=0.4\textwidth]{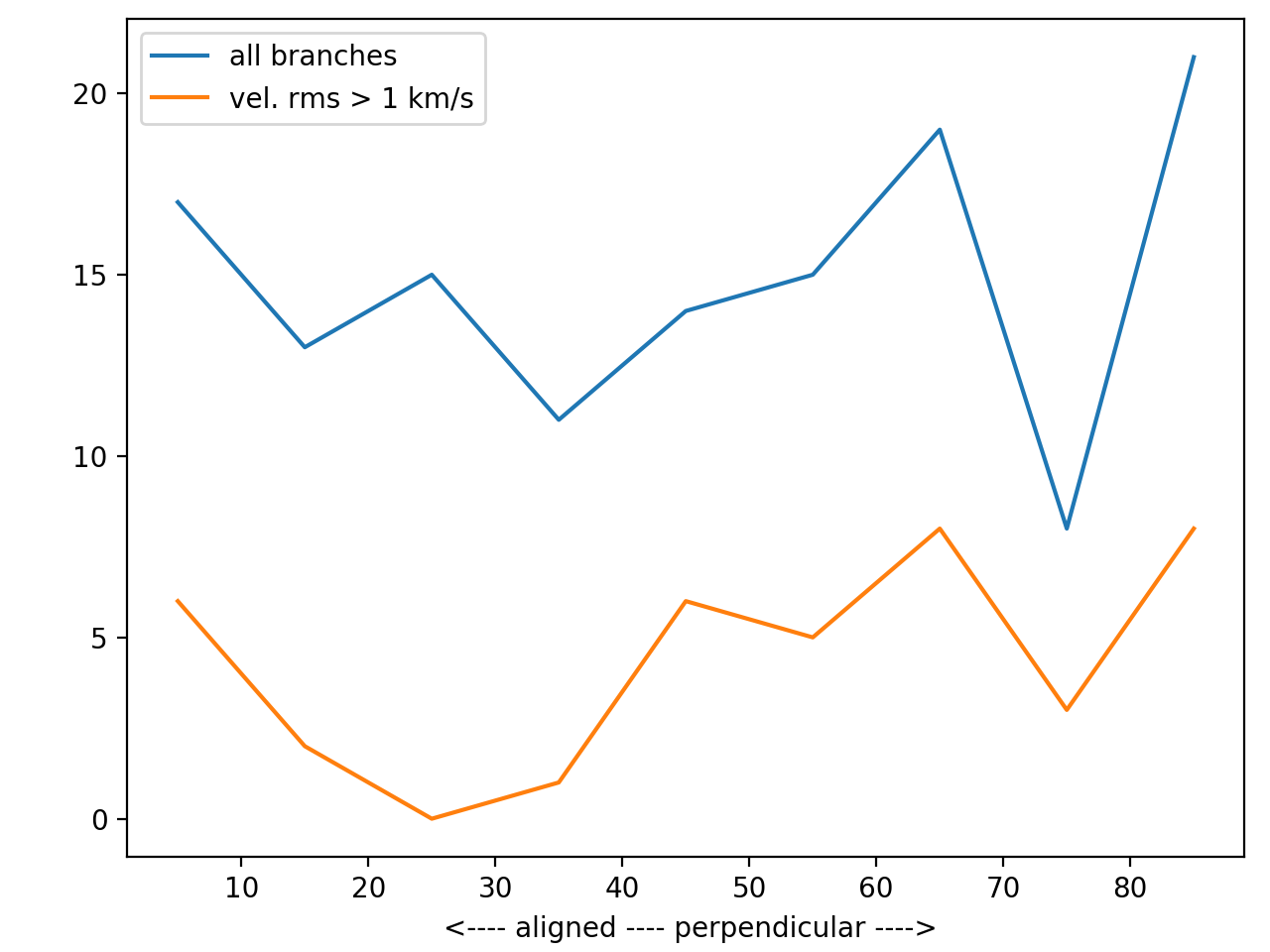}
  \caption{\label{velfield} (L) The intensity-weighted first moment map of \co{12}{2-1}, with filaments overlaid.  At each point, the local gradient in the 2-dimensional velocity field was measured.  (R) A histogram of the angles between each filament branch's skeleton and the local velocity gradient - an angle of zero would result of the velocity gradient were completely aligned with the filament branch.  Histograms are shown for all filament branches, and for those that are matched to elongated dendrogram structures.  No strong trend is evident in the alignment.}
\end{figure}

Part of understanding how stars accrete mass, and how that relates to molecular cloud structure, is determining the extent to which mass is accreted along filaments.  To begin to quantify filamentary accretion in this region, we calculated the intensity-weighted mean velocity at each point in the map (the first moment of the intensity cube), and then calculated the 2-dimensional gradient of that velocity field.  Figure~\ref{velfield}(L) shows the \CO{12}{2-1} moment 1 velocity map, with filaments overlaid.  One diagnostic is whether the velocity gradients are predominantly aligned with filaments, as would be the case if accretion along relatively long-lived filaments were dominant, or across the filaments, as would be the case if the filaments were the result of a turbulent velocity field, and not very dynamically important in the cloud. Figure~\ref{velfield}(R) shows the distribution of angles between the local gradient in the moment 1 velocity, and the filament skeleton direction. No strong trend is evident in the alignment, although there is a weak trend for filaments with a velocity rms larger than 1$\;$km$\;$s$^{-1}$ to have a velocity gradient more across than along the filament. This result suggests that accretion along filaments, although it may be present, does not dominate over the stochastic turbulent velocity motions in the cloud.

\subsection{Structure analysis result: filament stability}
\label{filstable}

The stability of cylindrical shapes has been calculated for numerous
cases including infinite homogeneous \citep{CF53,ostriker64a},
polytropic \citep{ostriker64b}, and magnetized
\citep{stodol63,tilley03}. Unmagnetized isothermal filaments with line
mass i.e. mass per unit length $M_l > 2\sigma^2/G$ are unstable to
gravitational collapse, where $\sigma$ is the velocity dispersion and
$G$ the gravitational constant.  Other geometries and equations of
state change the critical value by a factor of order unity.  We
calculate the critical line mass at each point of each filament, for
only thermal support using the CO excitation temperature
i.e. $\sigma^2 = kT_{ex}/\mu$ where $\mu$ is the mean molecular mass
2.36$m_H$, and for thermal and turbulent support $\sigma^2 =
kT_{ex}/\mu + \sigma_v^2$ where $\sigma_v$ is the second velocity
moment of the filament, calculated over the part of the cube within
5$\;$km$\;$s$^{-1}$ of the filament peak velocity.

% other filament collapse refs
% http://adsabs.harvard.edu/abs/2005MNRAS.356.1429S
% Holden et al 2009 PASP

\begin{figure}
    \centering
\begin{minipage}[b]{3.3in}
    \hspace{1cm}\resizebox{2.58in}{!}{ \includegraphics[trim=0mm 10mm 0mm 0mm, clip]{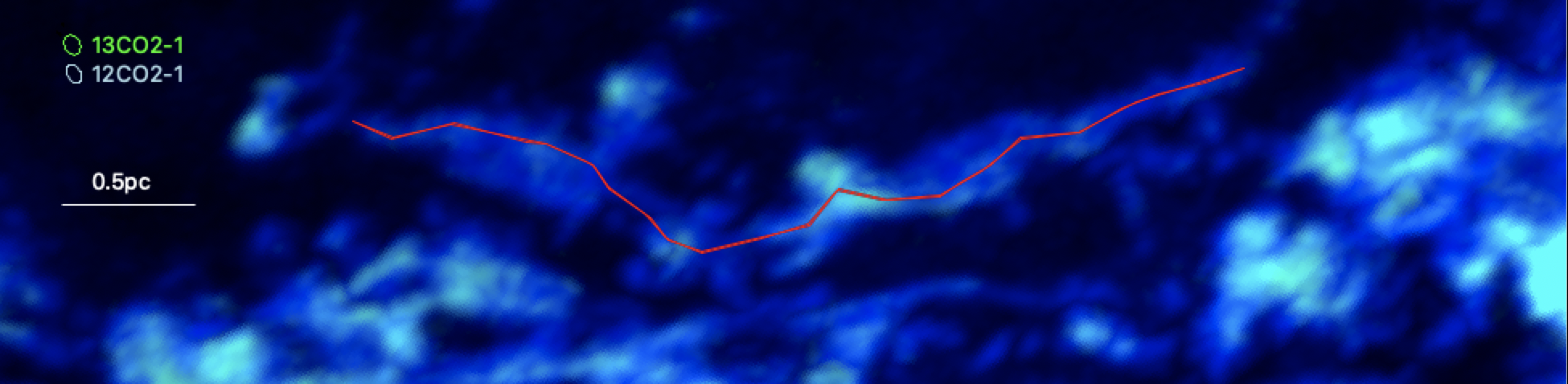}}\vspace{0em}
    \resizebox{3.3in}{!}{\includegraphics[trim=0mm 0mm 0mm 14mm, clip]{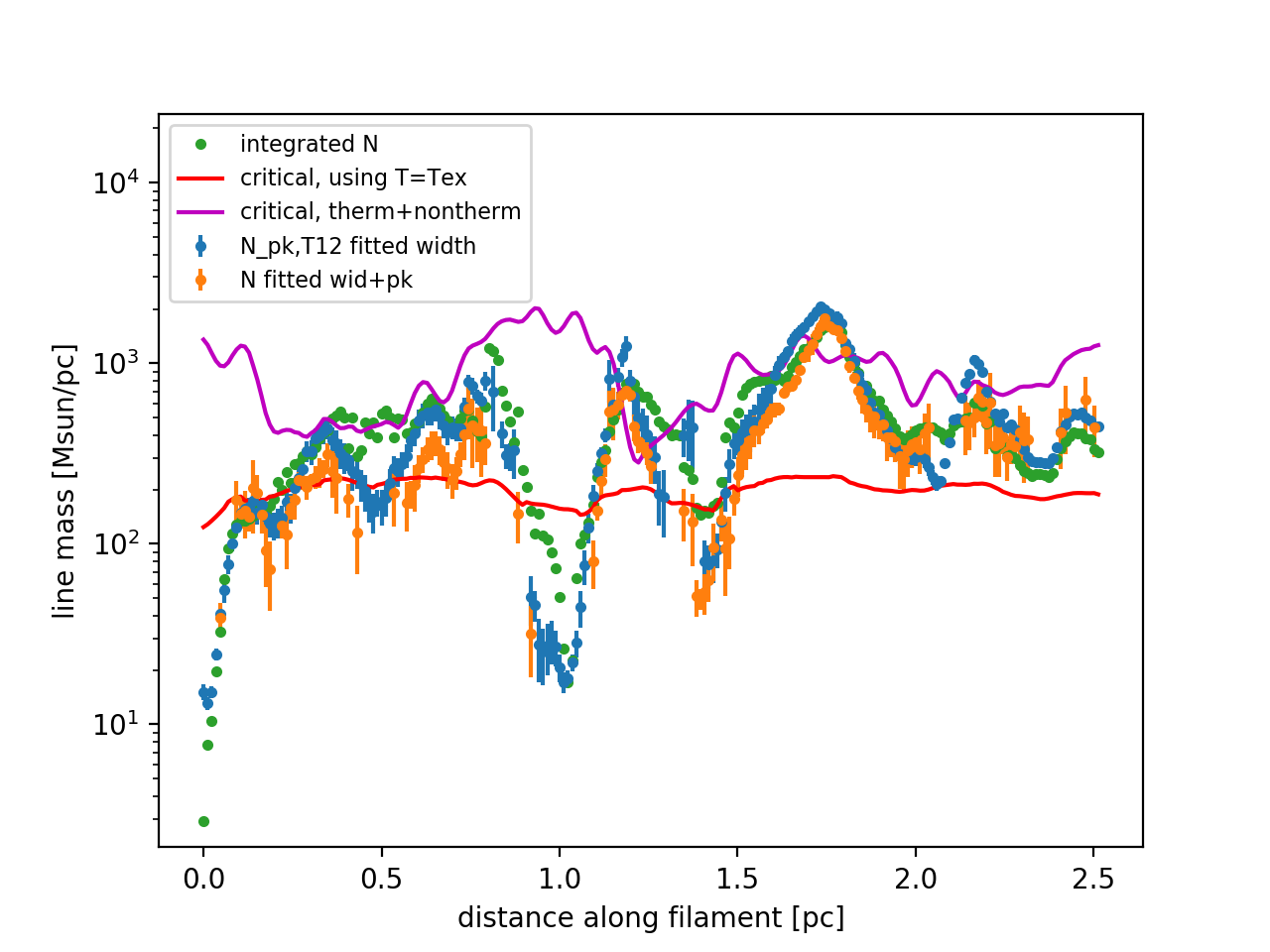}}
\end{minipage}

\caption{\label{onefil_fig} Top: A representative filament, showing $^{12}$CO and $^{13}$CO peak intensity as blue and green.  Bottom: Line mass calculated three different ways along the filament (green, blue, orange dots).   Critical mass for only thermal support (red), and thermal+nonthermal support (magenta), are also plotted.  The more visually fragmented portion of the filament $\sim$1pc along the filament (the ``gap'') has $M_l<<M_{crit}$. The bright knot $\sim$1.7pc along appears to be unsupported against collapse, with $M_l>M_{crit,nonthermal}$.}
\end{figure}

We calculate the actual line mass three ways: for each point on the
filament, {\tt filfinder} extracts the profile perpendicular to that
point.  We fit a Gaussian of width $\sigma_r$ to the profile of
\co{12}{2-1} intensity, and calculate $M_l =
\sigma_r\sqrt{2\pi}N_{pk}$ using the LTE column density in the center
of the filament at that point, $N_{pk}$. The fitted widths and peak
column densities are calculated from the average of a 3-pixel
neighborhood (smaller than the angular resolution) to increase
signal-to-noise.
In the second method, we fit the width and amplitude of the
perpendicular profile in $N$, and use $M_l = \sigma_r\sqrt{2\pi}N_{amp}$.
In the third method, we simply integrate $N$ across
each point of the filament.  When structures are close together in the direction perpendicular to the filament axis, the
width-fitting methods do not have as many points to fit
that are unambiguously associated with the given filament,
and thus can underestimate $\sigma_r$ and $M_l$, but on the other
hand, the integrated $N$ will be an over estimate because it includes
unassociated emission.  Figure~\ref{onefil_fig} shows the three line
masses and two critical masses along one example filament.
The more visually fragmented portion of the filament $\sim$1pc from the bottom (the ``gap'') has $M_l<<M_{l,crit}$ and that material is likely not gravitationally bound. By contrast, the bright knot $\sim$1.7pc from the bottom of the inset (from the left in the profile) appears to be unsupported against collapse, with $M_l>M_{crit,nonthermal}$.  Support for other portions of the filament are more ambiguous, with $M_{crit,thermal}<M_l<M_{crit,nonthermal}$.

\begin{figure}
  \includegraphics[width=0.45\textwidth]{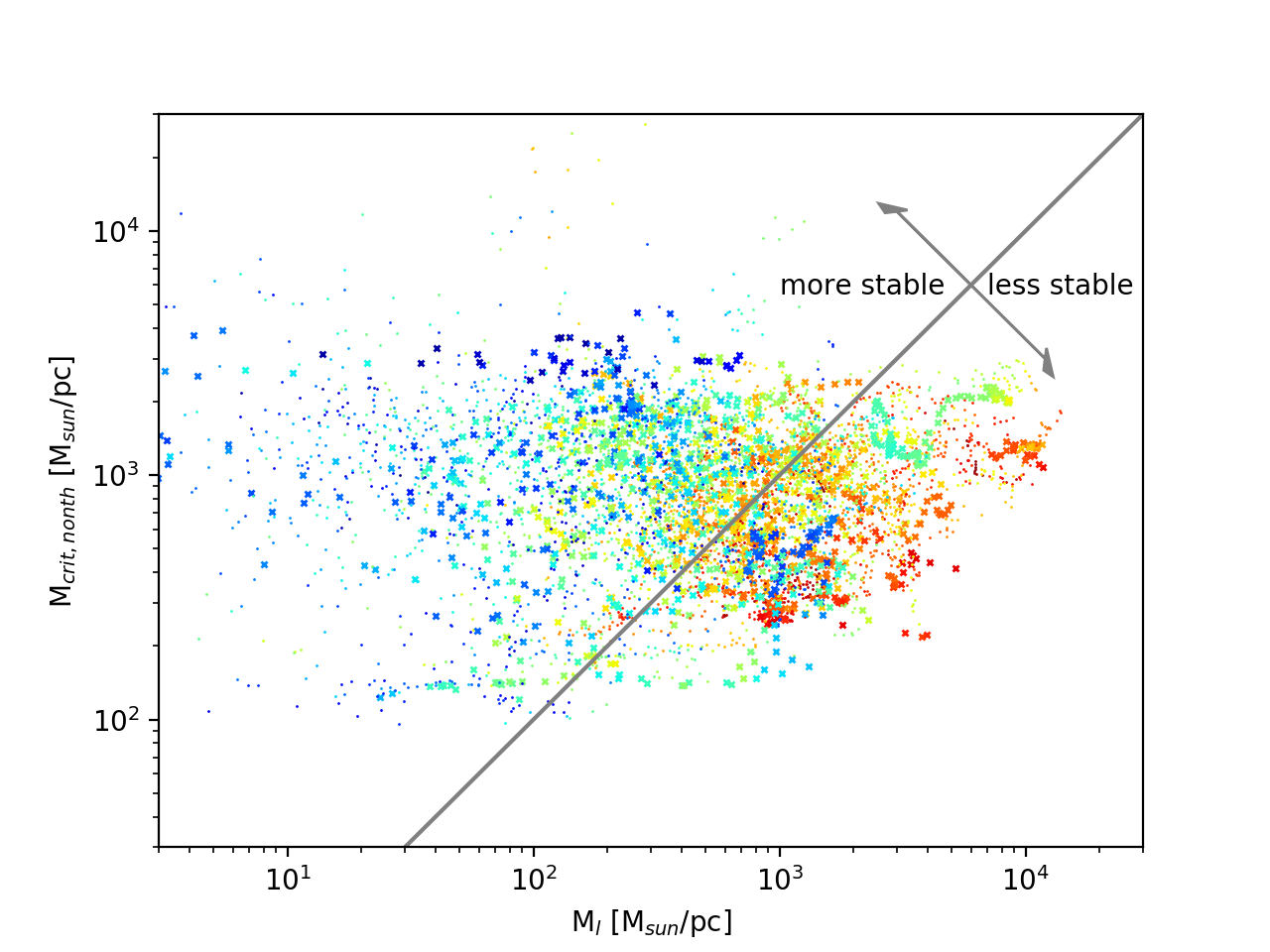}
  \includegraphics[width=0.45\textwidth]{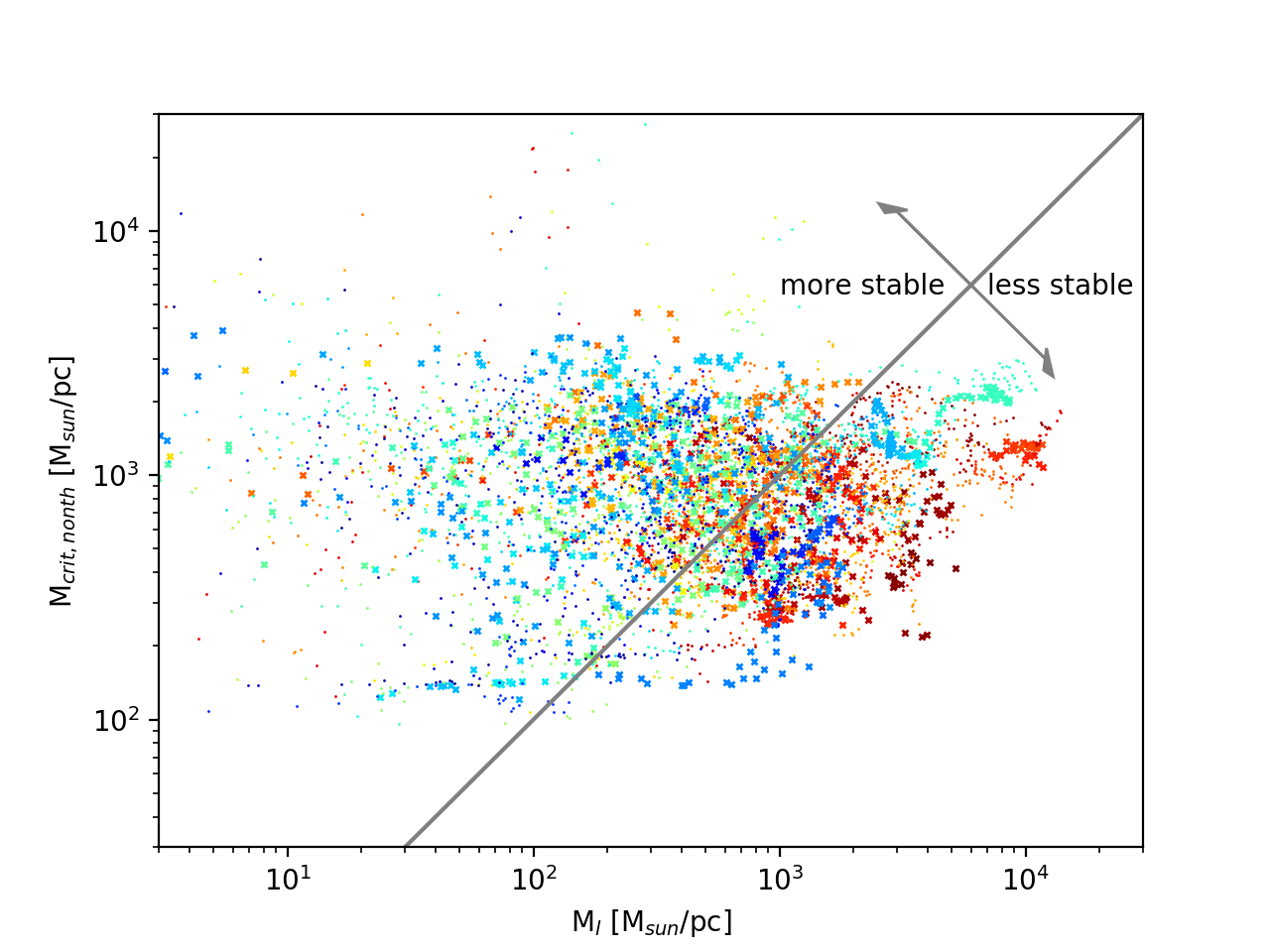}
  \caption{\label{filstability} Filament stability: points with $M_l>M_{crit}$ have more mass per unit length than thermal or kinetic support can prevent from collapsing. (L) points colored by \co{12}{2-1} brightness temperature from lower (blue) to higher (red).  ``x'' symbols are those filaments that correspond well to elongated isointensity contours (Figure~\ref{fils_dendro_fig}).  (R) points colored by 8$\mu$m brightness from lower (blue) to higher (red).}
\end{figure}

The gravitational stability of 0.1pc scale structures in 30Dor-10
analyzed as filaments is shown in Figure~\ref{filstability}. All points for all filaments are now plotted together, after analyzing all filaments as shown in Figure~\ref{onefil_fig}.
Points
with $M_l>M_{crit}$ have more mass per unit length than thermal or
kinetic support can prevent from collapsing (the nonthermal critical
mass is shown). The points cluster at $M_l=M_{crit}$, but there are
many with $M_l>M_{crit}$ which might be unstable, or supported by
magnetic fields.  There is a systematic uncertainty in $M_l$ depending
on the choice of abundance ratio $n(H_2)/n(^{13}CO)$.  5$\times$10$^6$
was used here; a lower ratio, or higher $^{13}CO$ abundance, by a
factor of 5 would move most points into the stable part of the plot.
Alternately, a factor of 3 lower ratio combined with a systematic
overestimate of $N_{LTE}$ by a factor of two (Figure~\ref{Nrecoveryfig}, \S\ref{Nrecoverysection})
would also bring most points into the stable regime. 
Given these
systematic uncertainties it may be more illuminating to examine the
trends - clearly, locations with brighter \co{12}{2-1}, or those with
brighter 8$\mu$m emission \citep[from {\it
    Spitzer}/SAGE][]{meixnersage}, are statistically less
gravitationally stable.  Both trends are present when all filament
branches are considered (dots) as well as when only considering
filaments that are associated with elongated isointensity contours
(crosses, see Figure~\ref{fils_dendro_fig} and \S\ref{fils_dendro_section}.  It is not unexpected that
less stable parts of the cloud might have more significant associated
star formation, and hence brighter 8$\mu$m emission, but the {\em
  Spitzer} resolution of $>$0.5pc makes it difficult to unambiguously
associate emission with 0.1pc scale molecular structures or to draw
strong conclusions.

\begin{figure}
  \includegraphics[width=0.45\textwidth]{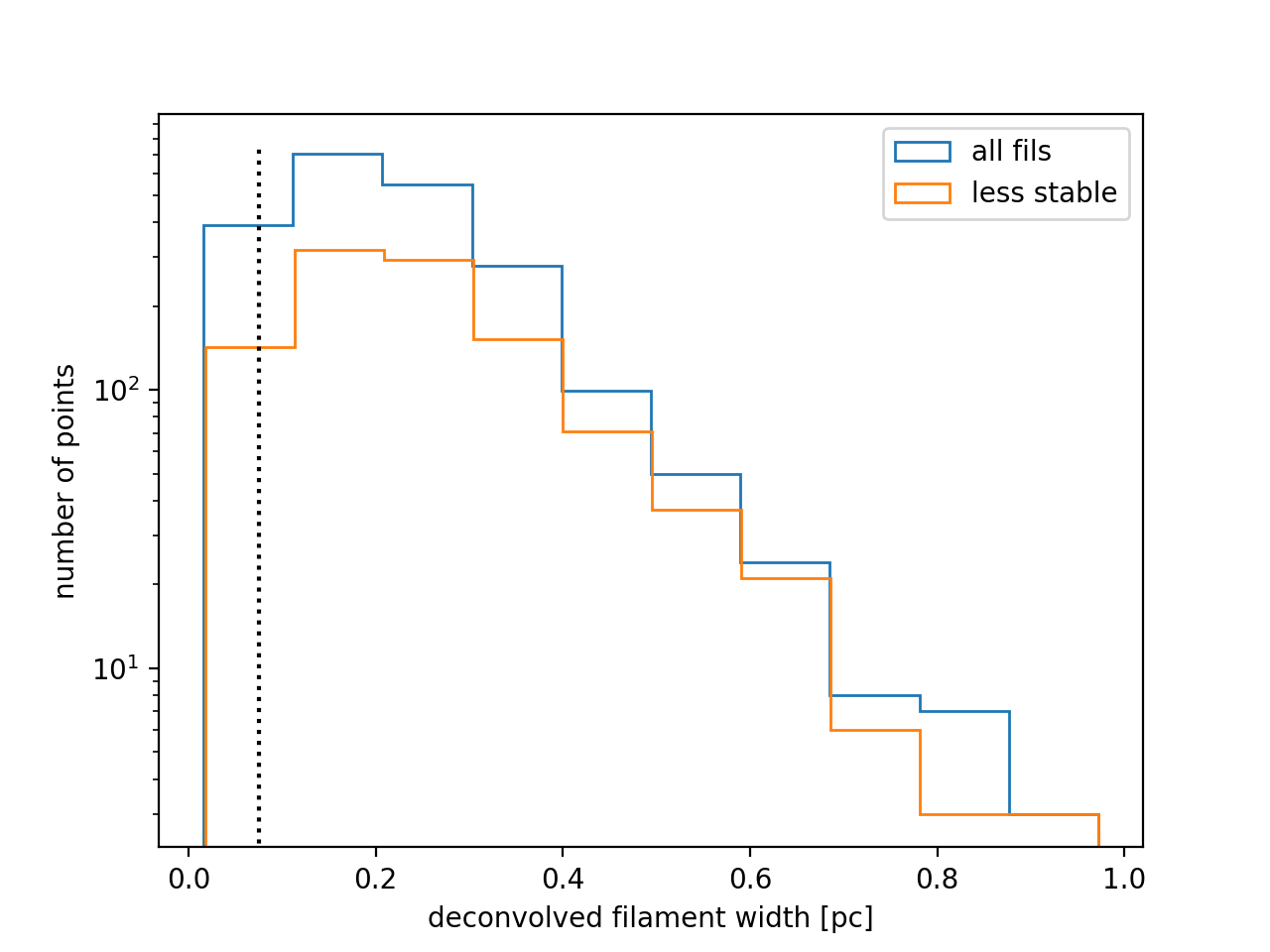}
  \caption{\label{filwidth} Distribution of widths of positions along filaments, where ``width'' = 2.534 times the fitted spatial moment of intensity, with the beam size of 0.075pc subtracted in quadrature.  All points with fitted width $>$ 2 times the width uncertainty are plotted, as well as the subset with $M_l > M_{crit,nonth}$. The dotted line indicates the beamsize.}
\end{figure}

In Figure~\ref{filwidth} we present the width distribution of filaments.
The spatial moment of $^{12}$CO intensity perpendicular to the filament was calculated at each position, as well as the spatial moment of calculated column density.  The weighted mean of those moments is used as the best estimator of the spatial sigma $\sigma_r$, and the deconvolved width is $\sqrt{(2.354\sigma_r)^2-0.075^2}$.  There is no significant difference in the width distribution between points on filaments that have line mass greater than critical line mass including nonthermal (turbulent) support.  The median width is 0.09pc, with standard deviation of 0.06pc, so the filaments in 30~Dor-10 have consistent widths to the 0.1pc width found in solar neighborhood clouds \citep{arzoumanian}.

%=================================================================================
\subsection{Structure analysis result: size-linewidth-flux relations}
\label{sizelinesection}

\begin{figure}[h]
  \includegraphics[width=0.45\textwidth]{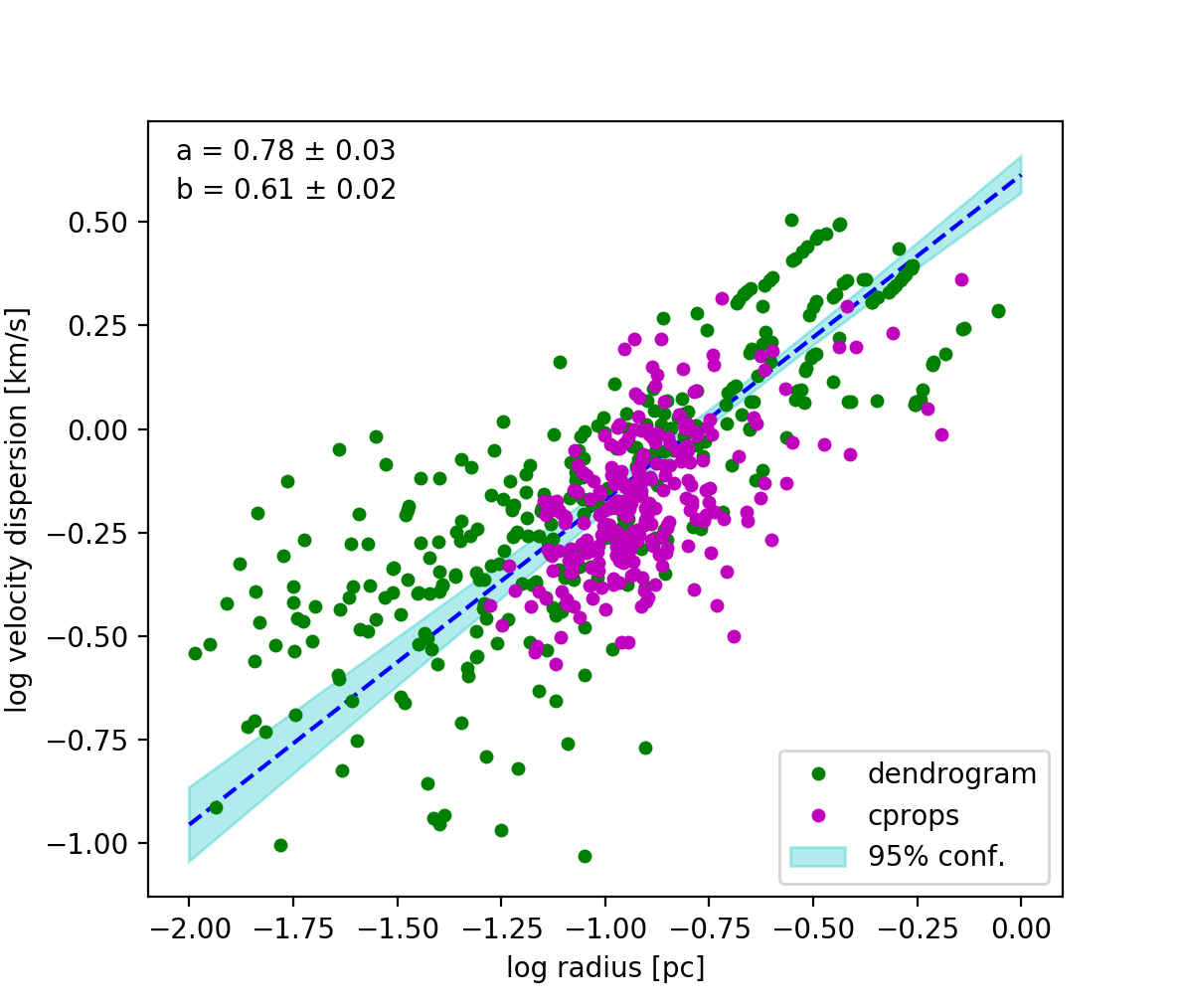}
  \includegraphics[width=0.45\textwidth]{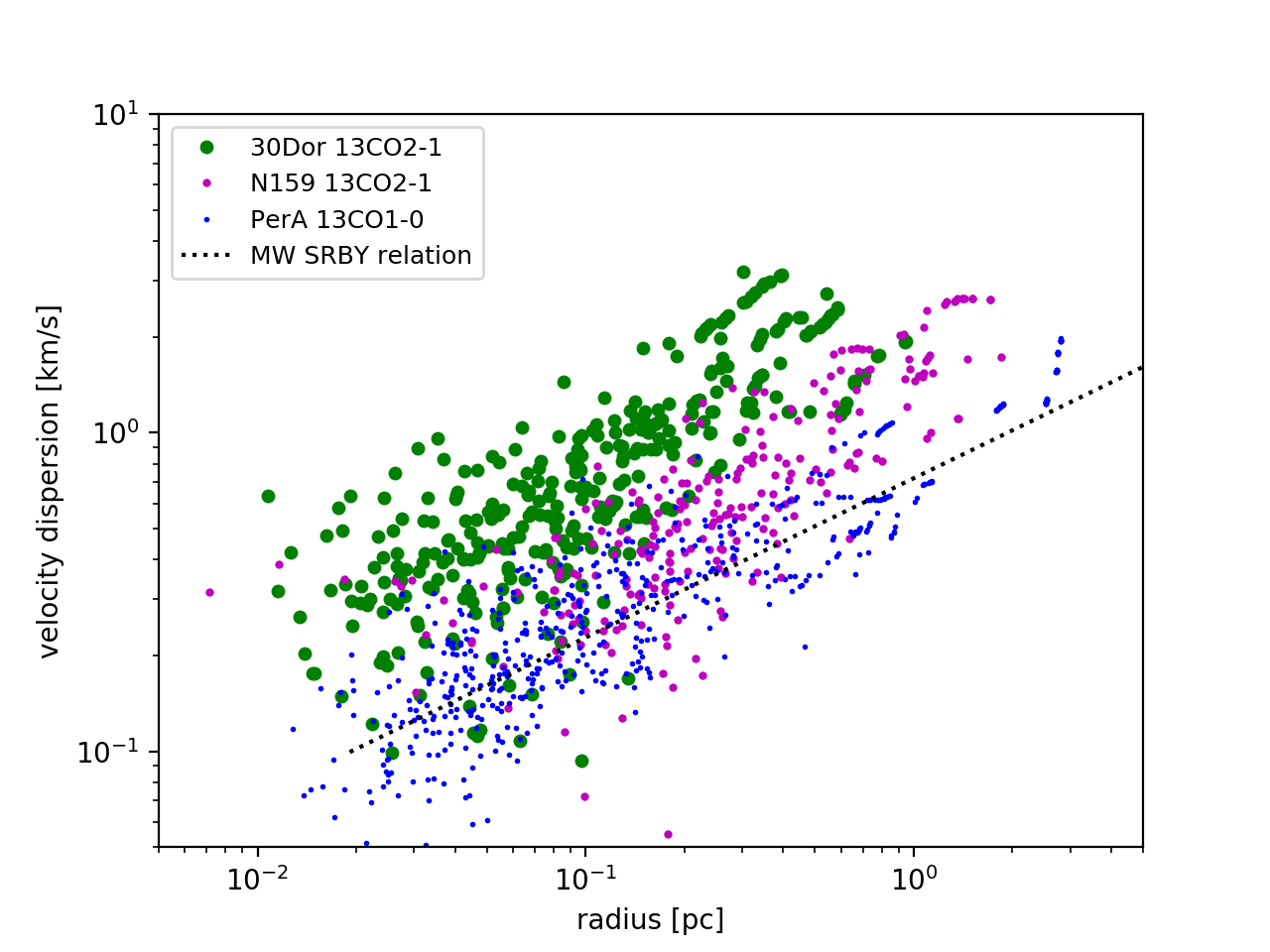}
  \caption{(L) Size-linewidth relation in 30Dor-10 for dendrogram (green) and clumpfind (magenta) structures.  Overlaid is the slope fitted to dendrogram structures $\log(\sigma_v)=a\log(r) + b$. (R) Comparison of size-linewidth relations between 30Dor-10, the nearby less evolved LMC massive star formation region N159, Galactic molecular cloud PerseusA, and the ``historic'' Milky Way disk relation \citep[dashed line; the data from which that was derived are all off the right-hand side of the plot][]{srby,heyer09}.  Data from all regions were re-analyzed the same way, with clumps segmented using dendrograms.}
    \label{sizelinefig}
\end{figure}

The relations between size, velocity dispersion, and surface density (or luminosity as a proxy for mass) have long been studied to yield insight into the structure and gravitational stability of molecular clouds \citep[e.g.][]{heyer09,goodman09,ballesteros10,bolatto}.  Figure~\ref{sizelinefig} shows the relation for this cloud, for structures segmented with {\tt dendrograms} and {\tt clumpfind}, and compared to other molecular clouds.
The ``radius'' plotted is the historical value of 1.91 times the second moment in the plane of the sky, and both that size and the velocity dispersion have had the instrumental resolution deconvolved.
The differences between algorithms are clear - the dendrogram analysis classifies all emission, down to very small structures, and multiply-counts that emission, as part of very large structures. By contrast, clumpfind is highly biased towards selecting clumps in a narrow range of sizes.
% TODO MARK VEL, BEAM ON PLOTS?
The relation is fit to a power-law with the Levenberg-Marquardt least-squares fitter in the Kapteyn package \citep{kapteyn}.
The slope of 0.75$\pm$0.05 is somewhat steeper than the theoretical slope of 0.5 for a medium dominated by turbulent shocks \citep[][and refs therein]{ballesteros10},
and on the higher side but within 3$\sigma$ of values measured in other clouds \citep[see references and discussion of 30Dor-10 at lower angular resolution in ][]{nayak16}.

N159, a pair of massive star forming clouds just to the south of 30~Doradus, has been observed with ALMA at similar angular resolution to the data presented here (\citet{saigo17}: resolution of 1.21\,\arcsec$\times$0.84\,\arcsec, \citet{tokuda18}: resolution of 0.29\,\arcsec$\times$0.25\,\arcsec, \citet{fukui15}:resolution of 1.3\,\arcsec$\times$0.8\,\arcsec, \citet{fukui18}:  resolution of 0.28\,\arcsec$\times$0.25\,\arcsec)
dendrogram structures in N159 have a size-linewidth slope of 0.7$\pm$0.1, consistent with 30Dor-10, but the linewidth at a given scale is 0.33$\pm$0.05 dex lower.  The molecular mass of N159 is significantly higher than 30Dor-10, so the enhanced linewidths in 30Dor cannot be explained by global equilibrium between gravitational and kinetic energy.  The enhanced kinetic energy is either a result of feedback, or an equilibrium including external pressure on the molecular cloud.  There is embedded star formation within 30Dor-10, but the total infrared luminosity and inferred star formation rate is again significantly less than in N159.
Thus, internal sources of mechanical energy are unlikely the source of elevated turbulence,
and instead this is either caused by thermal pressure from the ionized gas, or large-scale dynamics of the region.
Interestingly, \citet{lee_30dor} analyze high-J CO and far-infrared line emission at low spatial resolution in 30~Doradus, and find that they require significant energy input from low-velocity shocks to explain the CO line spectral energy distribution.  They also conclude based on the distribution of main sequence and protostellar wind sources that local kinetic energy injection is unlikely to dominate, but instead they favor kpc-scale energy injection related to the overall dynamics of the region.   This kpc-scale energy input is very likely related to kpc-scale colliding filaments which intersect at exactly the center of 30~Doradus, and may well be the reason that the super-star cluster could form there in the first place \citep{fukui_30dor_formation}.

Also shown in Figure~\ref{sizelinefig} are the relation fit to Galactic molecular clouds $\delta v$=0.72$\;$R$^{0.5}$ \citep[][SRBY]{heyer09,srby}, and structures in the Perseus$\;$A cloud in the solar neighborhood \citep{ridge06}.  The fitted slope of 0.5$\pm$0.05 and intercept are consistent with the SRBY relation, with dispersion at a given scale 0.65$\pm$0.05 lower than in 30Dor-10.   
% dor, n159, per offsets = 0.61,0.27,-0.06

\begin{figure}
  \includegraphics[width=0.45\textwidth]{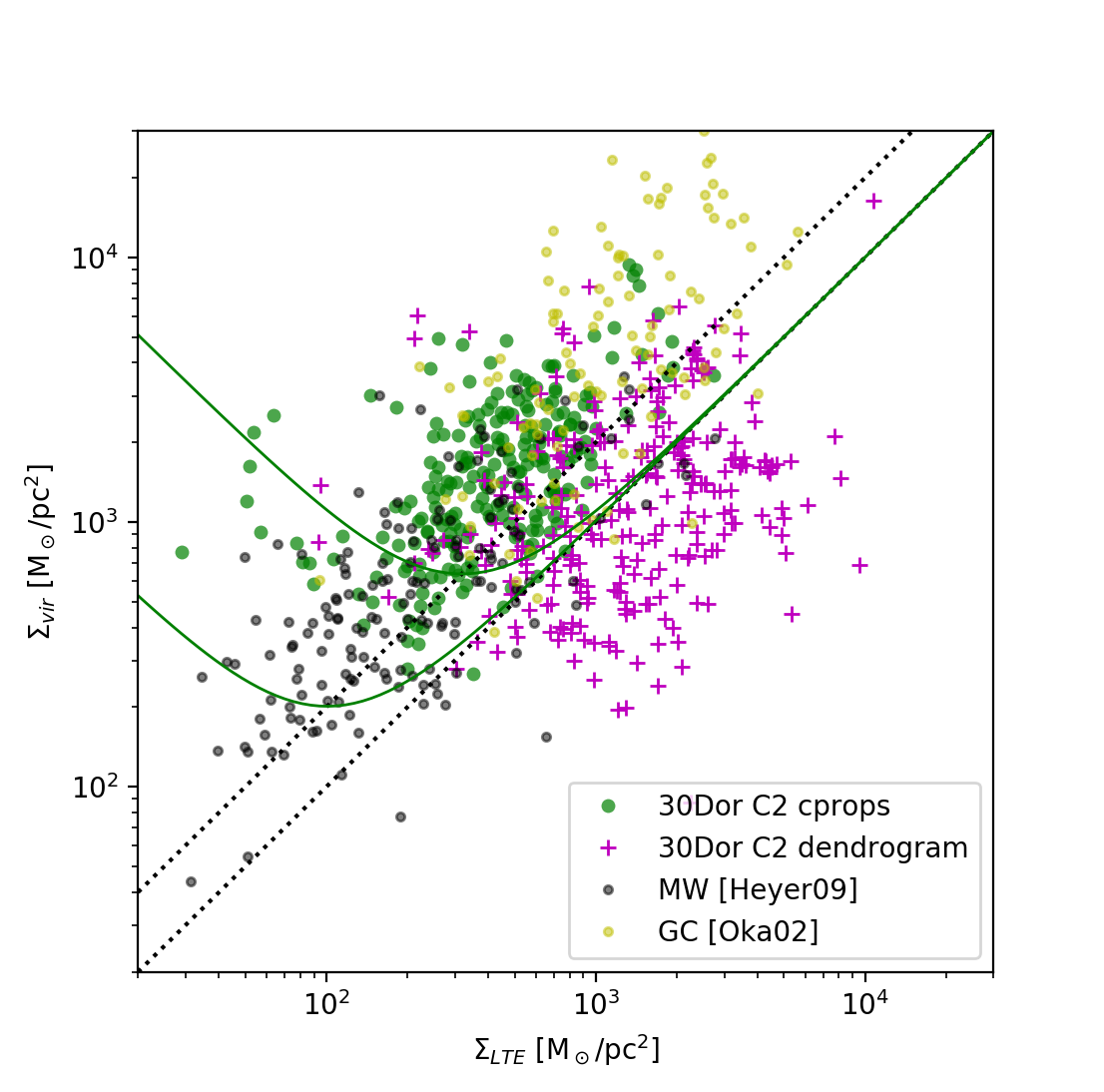}
  \includegraphics[width=0.45\textwidth]{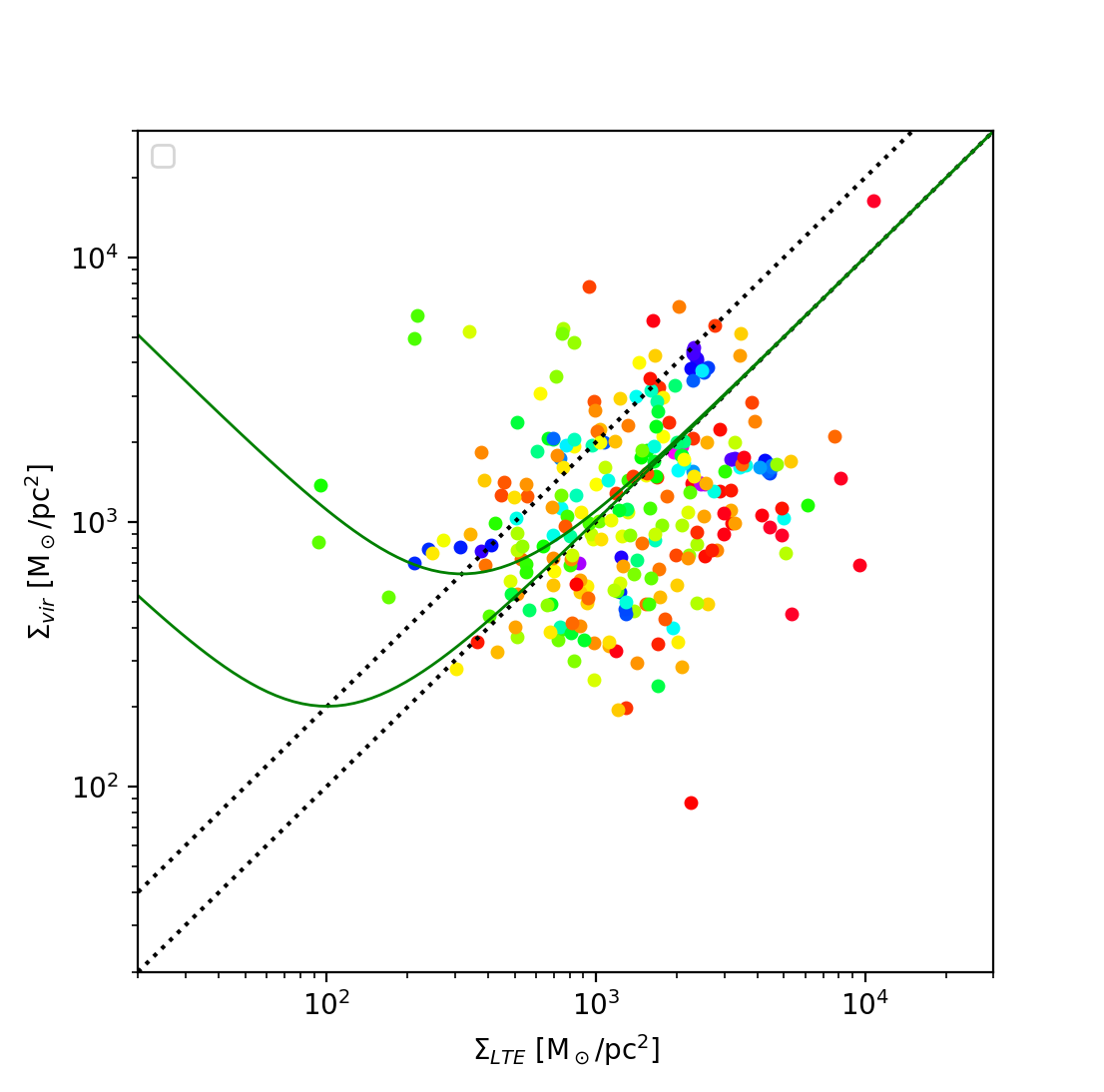}
  \caption{\label{boundfig} Relation between virial surface density 5$\sigma_v^2$/(G$\pi$R) and LTE surface density M$_{LTE}$/($\pi$R$^2$).  Virial equilibrium between gravitational and kinetic energy $\Sigma_{vir}$=$\Sigma_{LTE}$ and gravity-driven collapse $\Sigma_{vir}$=2$\Sigma_{LTE}$ are marked as dashed lines.  Virial equilibrium with external pressures $nT$=(10$^5$,10$^6$)cm$^{-3}$K are solid green curves. (L) the different segmentation methods in this region (green=cprops, magenta=dendrogram) are compared to previous measurements of the galactic center \citep[CG;][]{oka01} and Milky Way disk \citep[MW;][]{heyer09}.
    (R) Only dendrogram structures in this region are shown, now colored by size from smaller=blue to larger=red.}
\end{figure}

Figure~\ref{boundfig} shows the relation between LTE and virial surface density for structures in 30Dor-10.  $\Sigma$=M/$\pi$R$^2$ where $R$ is the cloud ``radius'' 1.91$\sqrt{\sigma_x\sigma_y}$ calculated from the weighted spatial moments $\sigma_x$ and $\sigma_y$.  One can also use the ``exact area'' or full spatial extent of the pixels assigned to each structure - this area is 1.2-1.9 times larger, decreasing both surface densities accordingly.  $\Sigma_{LTE}$ is calculated from the LTE mass, and $\Sigma_{vir}$ from the virial mass 5$\sigma_v^2$R/G.  This plot is sometimes referred to as a ``boundedness'' plot because it probes the degree to which kinetic energy is gravitationally bounded.  Structures in virial equilibrium between gravitational and kinetic energy would lie along $\Sigma_{vir}$=$\Sigma_{LTE}$, and those in free-fall (hierarchical) collapse driven by gravity along $\Sigma_{vir}$=2$\Sigma_{LTE}$ \citep{ballesteros10}.  Alternately clumps might be in virial equilibrium between kinetic and gravitational energy with external pressure \citep{field10}, if they are sufficiently long-lived to achieve that virialization \citep{bonnell06}.  The resulting curves
\[ \Sigma_{vir} = \Sigma_{LTE} + {{20}\over{3\pi 20.9}}{{nT}\over{\Sigma_{LTE}}} \]
are marked on Figure~\ref{boundfig} for external pressures of $nT$= (10$^5$ , 10$^6$) cm$^{-3}$K.

Structures segmented using {\tt clumpfind} fall in a similar locus to much larger (1-10pc) molecular clouds in the Milky Way disk and Galactic central molecular zone \citep{heyer09,oka01}.  These all fall somewhat above the unity relation expected from virial equilibrium, and more consistent with either external pressure confinement or free-fall collapse.  Structures in 30Dor-10 lie in between the disk and central molecular zone (CMZ) clouds;  if the relevant model is confinement by external pressure $P_e$, then that pressure is higher in 30Dor-10 than typical Milky Way, but lower than the CMZ.  For the large Milky Way clouds, $P_e$ should be interpreted as pressure from the neutral interstellar medium surrounding the molecular cloud, but for the sub-parsec structures plotted in 30Dor-10, $P_e$ is the pressure external to a core, which is a combination of any warm ISM pressure acting on the entire cloud, and the effective pressure of the diffuse cloud acting on the core.  In Paper~I we used the formula of \citet{bertoldi92} to estimate that interclump pressure to be 3$\times$10$^6$cm$^{-3}$K, consistent with the location of core and clumps in Figure~\ref{boundfig}.

Structures identified by {\tt dendrogram} have significantly higher $\Sigma_{LTE}$ than those identified with {\tt clumpfind}.  This conclusion is unchanged if either the ``exact'' cloud area is used to calculate $\Sigma$, or if observed quantities (without the velocity and angular resolution deconvolved) are plotted.
Dendrogram structure properties are calculated in Figure~\ref{sizelinefig} by assuming all emission above each isointensity surface is part of the structure (bijection), but one could also subtract the value of that lower bounding surface (clipping), as discussed in \citet{cprops}.
Clipping decreases $\Sigma_{vir}$ by only 0.02dex, whilst decreasing $\Sigma_{LTE}$ by 0.3dex - the mean difference of $\Sigma_{LTE}$ between {\tt dendrogram} and {\tt clumpfind} structures is 0.8dex without clipping, and decreases to 0.5dex when clipping is used.  \citet{cprops} conclude that clipping underestimates clump masses for their solar neighborhood clouds, but for the data presented here, clipping would result in structures with $\Sigma_{LTE}$ closer to those identified by {\tt clumpfind}.

There is some tendency for larger dendrogram structures to be evidently less gravitationally stable: The second panel of Figure~\ref{boundfig} shows $\Sigma_{vir}$ and $\Sigma_{LTE}$ colored by structure size.  The virial mass was calculated for a constant density sphere, but molecular cloud substructures are clearly not spheres.
Generalizing from spheres, the gravitational energy of an ellipsoid can be written in closed form for various radial density profiles \citep{neutsch79}, including constant density,
\[ U = {3\over 5}{{GM^2}\over l}w(\eta)\]
\[ w(\eta) = {{\eta\sinh^{-1}\left(\sqrt{\eta^2-1}\right)}\over\sqrt{\eta^2-1}}, \]
where $M$ is the mass, $l$ the semi-major axis, and $\eta = l/R$ the
aspect ratio \citep{lee17}.  For the same mass, a more elongated cloud has greater gravitational energy by a factor of (2,3) for a cloud aspect ratio of $\sim$(3.5,10).  Thus the virial mass required for gravitational energy to balance a given kinetic energy is lower by a factor of a few for an elongated cloud compared to the assumed spherical shape.
If we applied this correction, we would conclude that the clouds are even less gravitationally stable, and located even lower down in Figure~\ref{boundfig}.   We measured the elongation or aspect ratio of each structure in the Figure, but there is no clear trend with boundedness, and applying an elongation correction does not decrease the scatter of points in the plot.

Another source of systematic uncertainty was mentioned in \S\ref{filstable}:
A factor of 5 higher assumed $^{13}CO$ abundance 
would move most dendrogram points into the stable part of the plot.
Alternately, a factor of 3 lower ratio combined with a systematic
overestimate of $N_{LTE}$ by a factor of two (Figure~\ref{Nrecoveryfig})
would also bring most points into the stable regime.  However, either of those corrections would cause the {\tt clumpfind} clumps to disagree significantly with previous measurements of molecular clouds in this and many other regions.

\begin{figure}
  \includegraphics[width=0.45\textwidth]{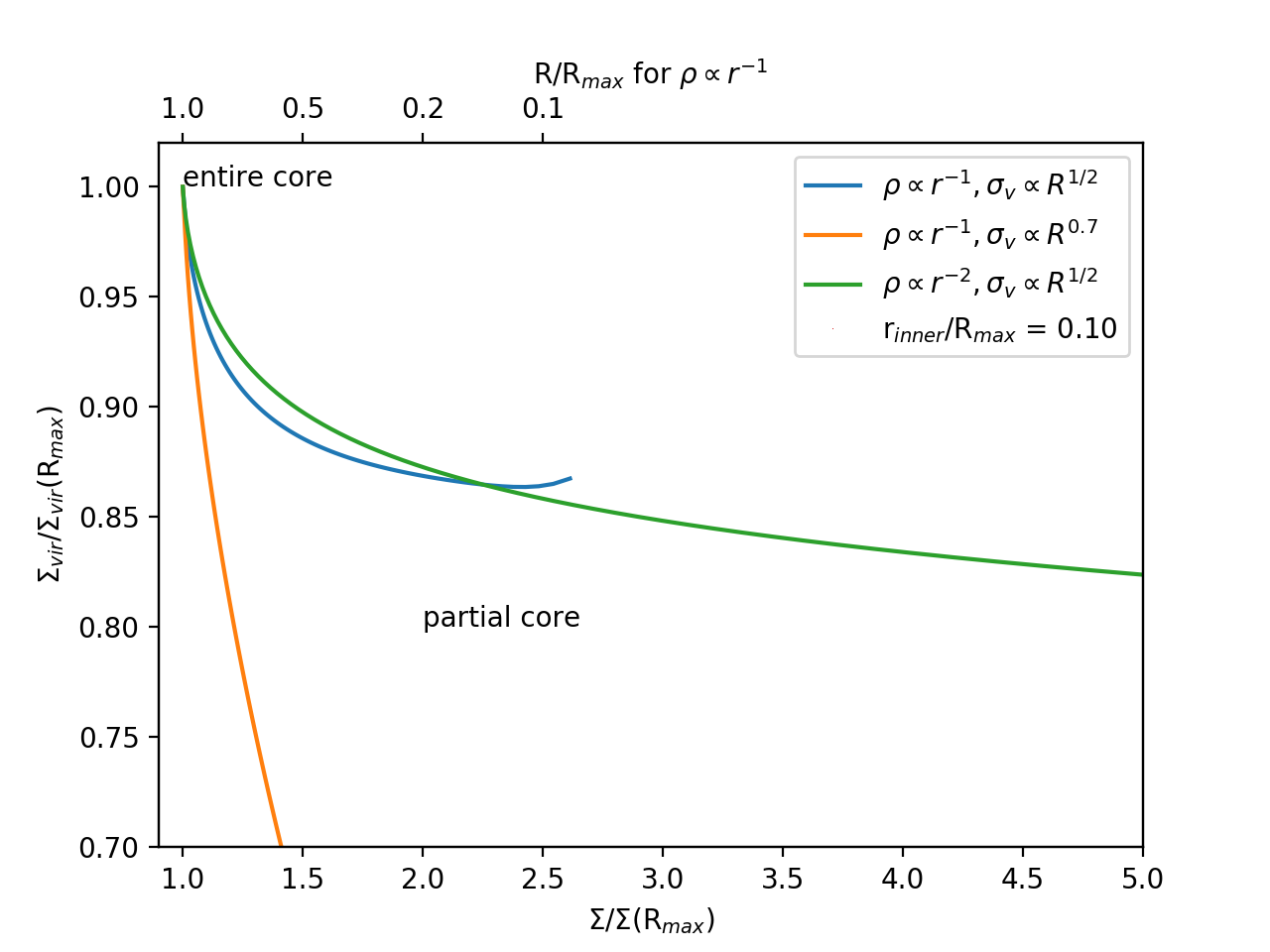}
  \caption{\label{SigSigVir} Relation between virial surface density 5$\sigma_v^2$/(G$\pi$R) and LTE surface density M$_{LTE}$/($\pi$R$^2$) for a model spherical core.  The core has constant density within r$_{inner}$, and a power-law radial density distribution outside of that.
    Each part of the core within a projected radius $R$ is assumed to follow the size-linewidth relation $\sigma_v\propto R^\alpha$.  Calculation of the two surface densities for a partial core will yield significantly higher $\Sigma$ and moderately lower $\Sigma_{vir}$ compared to the values calculated for the entire core.}
\end{figure}

The most likely cause for the difference between the boundedness of {\tt clumpfind} and {\tt dendrogram} segmented structures is that the {\tt dendrogram} structures by design do not include entire clumps.  Consider a model core with power-law density distribution outside of a core radius $r_c$.  The density distribution $\rho\propto r^{-1}$ for example was used in \citet{srby} to derive the commonly used relation between the second spatial moment $\sigma_x$ and the ``edge'' or ``effective radius'' of the core $R_e$ = 1.91 $\sigma_x$.  If one assumes that the velocity dispersion $\sigma_v$ follows a power-law size-linewidth relation $\sigma_v \propto R^\alpha$, where $R$ is the projected radius in the plane of the sky, then one can calculate the virial surface density
\[ \Sigma_{vir}(R) = { {5 \sigma_v^2}\over{\pi G M(R) R_{measured}} } \]
as a function of projected radius $R$, where $M(R)=\int_0^R4r\rho(r)dr$.
The measured radius $R_{measured}$ is what one could measure from the data within a radius $R$, and could be simply $R$ itself (i.e. the radius of the assignment area $\sqrt{{\rm area}/\pi}$), or more commonly the ``Solomon effective radius'' calculated from the two spatial second moments of the emission, 1.91 $\sqrt{\sigma_x \sigma_y}$.  Figure~\ref{SigSigVir} shows the virial surface density $\Sigma_{vir}$ and the actual measured surface density $\Sigma = M/\pi/R_{measured}^2$, as a function of $R$ for a model core with two different size-linewidth relations $\alpha$ and two different power-law density profiles.  Clearly, if one calculates a surface density for only the brightest part of a core, one will measure a significantly higher surface density $\Sigma$ than for the entire core.  At the same time, the calculated virial surface density $\Sigma_{vir}$ will be somewhat lower than the whole-core value.  This underscores the need to use the same segmentation method when comparing different datasets.
It also suggests that if parts of clouds segmented with dendrograms are analyzed,
the relative boundedness between two different clouds is a robust comparison, but the absolute value of that boundedness relative to the theoretical lines of stability should be interpreted cautiously.
As a final test, we calculated the size-linewidth-mass relations for structures identified with {\tt clumpfind} but using a higher and higher noise floor or cutoff, raising it gradually up to 20$\times$ the noise level in the cube.
Raising the cutoff floor causes the assigned structures' properties to smoothly move over to the location of the {\tt dendrogram}-assigned structures, as expected.

\subsection{Structure analysis result: Core mass function}

A fundamental question in star formation is whether the stellar
initial mass function (IMF) is predetermined by the molecular cloud
structure.  Similarity between the dense core mass function and the
IMF would support that premise.  We analyze compact structures
identified using {\tt clumpfind} (as implemented in {\tt cprops}).
The distinction between calling a
molecular cloud structure a clump or core is inhomogeneous in the
literature, with $\sim$1pc structures usually called clumps and
$\sim$0.1pc structures usually called cores; These structures are
0.1-0.2pc in diameter, so we call them cores for brevity, without
intending any implication about stability or concentration.

\begin{figure}
  \includegraphics[width=0.45\textwidth]{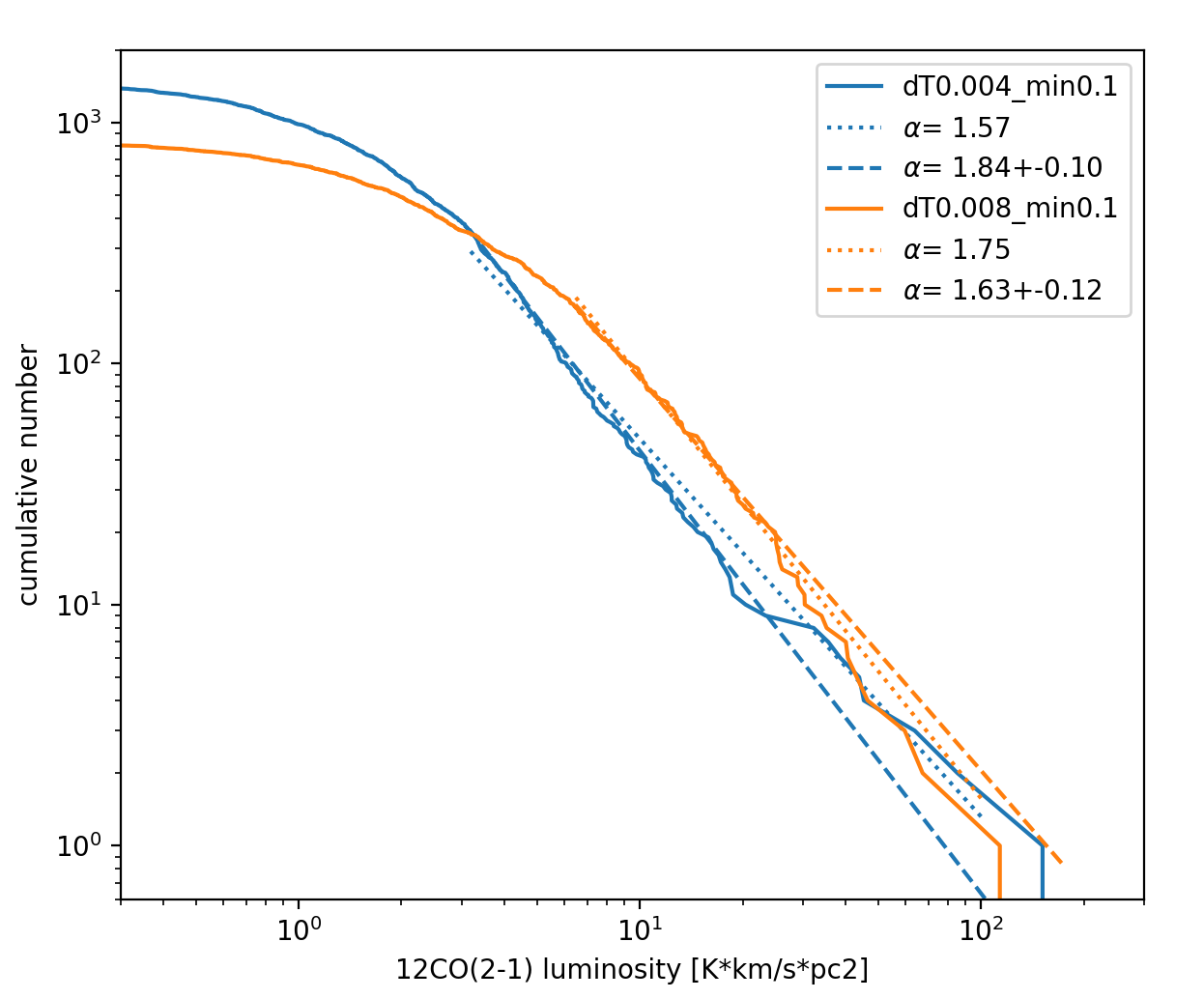}
  \includegraphics[width=0.45\textwidth]{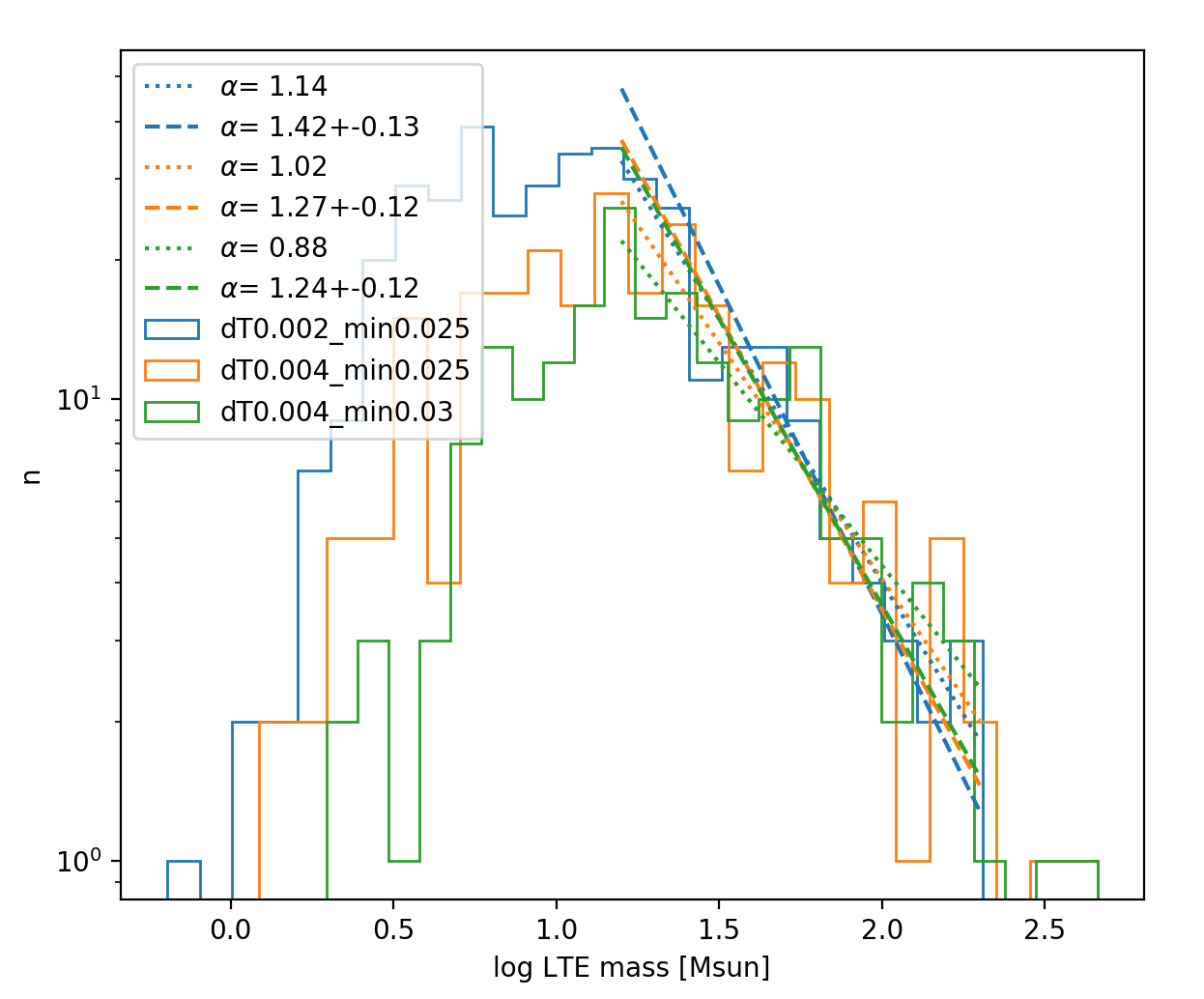}
  \caption{\label{intensity_functions} (L) cumulative luminosity distribution of \co{12}{2-1} {\tt clumpfind}-identified cores with two different sets of segmentation parameters.  Power-law fits are shown using a maximum likelihood estimator \citep[][eq 3.1; dashed]{fitlumfunc} and a simple linear fit to the log number (dotted). (R) differential LTE mass function of cores with three different segmentation parameters.}
\end{figure}

Figure~\ref{intensity_functions} shows the core luminosity and mass
distributions.  Different values of the {\tt clumpfind} segmentation
parameters are shown -- {\tt dT} and {\tt T$_{min}$}, the difference
between each local maximum and nearest saddle, and minimum level to
analyze.  The main effect of varying those parameters is to increase
the number of faint cores when {\tt T$_{min}$} is decreased.  Fitting
power law distributions has well-studied uncertainties
\citep[e.g.][]{fitlumfunc,jesus}; we show both the Hill maximum likelihood
estimator with its statistical uncertainty (dashed lines), as well as
a simple linear fit to the log number $n$, weighted by
$n^{-0.3}$ (dotted lines; the exponent of the weight makes little
difference to the result).  It is evident that even with the
mathematical uncertainties due to fit method,  the uncertainties due to
how the emission is segmented into structures are even larger.

\CO{12} luminosity is proportional to cloud mass on large
($>$pc) scales \citep[e.g.][]{bolatto}.  Although one would expect the
luminosity of an optically thick line to be a less reliable mass
tracer on small scales when emission fills the beam, it is still
interesting to consider the shape of the \CO{12} luminosity
function (Figure~\ref{intensity_functions} left).  The bright end is fit with a power law of slope
$\alpha$=1.7$\pm$0.2.
If luminosity were proportional to mass, this would imply a core
differential mass function $N(>M)\propto M^{-\alpha}$ slope of somewhat
steeper than the stellar initial mass function (IMF) slope -1.35.
However, the \CO{12} optical depth increases systematically with
mass $\tau\sim$M$_{LTE}^{0.3}$, so the observed \CO{12} luminosity
function is expected to be steeper than the mass function by a slope of about 0.3,
and the inferred mass function from \CO{12} alone would then be consistent with
the IMF, within uncertainties.
The second plot of the figure shows the differential LTE mass
distribution (from \CO{12} and \CO{13}), for different core segmentation parameters.  Slopes are
fitted directly to the differential mass distribution as well as using
the aforementioned Hill maximum likelihood estimator.  The systematic
biases of direct fitting are more evident than when fitting the
cumulative distribution.  The best estimate slope of -1.3$\pm$0.15 is
consistent with the stellar IMF slope -1.35.

The upper end of the ``core'' mass distribution is consistent with a power-law with the same slope as the stellar initial mass function.  A shallower mass function (more massive stars) has been found for the main-sequence massive stars in 30Doradus \citep{science}.  The next generation of star formation in 30Dor-10 will either not have that massive stellar excess, or the stellar mass distribution is not predetermined by the current core mass function.
The discussion of stability above raises the natural question of whether the clumps and cores whose mass distribution is being analyzed {\it should} have any correspondence with the stellar mass function, especially if many of those cores are not gravitationally bound or collapsing.  To test this, we analyzed the distribution of only cores with $\Sigma_{virial}/\Sigma$ less than a threshold (we tried thresholds between 2 and 4; see Figure~\ref{boundfig} and discussion in section \ref{sizelinesection} about the normalization of $\Sigma_{virial}/\Sigma$).  Interestingly, the shape of the mass distribution, and the fitted slope of the upper end are {\it the same} for the more gravitationally bound subset(s) of cores.  If the mass distribution of cores is independent of core stability, and is related to the stellar IMF, then this result supports the notion that the core and cloud structure predetermines the stellar IMF, before gravitational forces become dominant.

%The most massive clumps are shown in Figure~\ref{mostmassive}

%=======================================================================================
\section{Conclusions}

High-resolution ($<$0.1pc) observations of the 30Dor-10 molecular
cloud 15$\;$pc north of R136 are presented.  The CO emission
morphology contains clumps near the locations of known mid-infrared
massive protostars, as well as a series of parsec-long filaments
oriented almost directly towards R136, most of which don't show signs
of embedded star formation.  The aligned filaments could possibly be ``pillars'' left behind by photoionization, or could be a relic of the initial collapse and formation of 30Doradus and the R136 star cluster.

Analysis of the cloud substructures is performed by segmenting emission into disjoint approximately round ``cores'' using {\tt clumpfind}, by considering the hierarchical structures defined by isointensity contours using {\tt dendrograms}, and by segmenting into disjoint long thin ``filaments'' using {\tt Filfinder}.  The balance between gravitational and kinetic energy of ``cores'' and dendrogram branches are analyzed with formulae appropriate to spheres and ellipsoids, and the balance is analyzed for filaments using formulae appropriate to infinite cylinders.  We find that the filaments have widths of $\sim$0.1pc, similar to those in solar neighborhood clouds.

There is elevated kinetic energy (linewidths at a given size scale) in 30Dor-10 compared to other LMC and Galactic star formation regions.  The slope of the size-linewidth relation is also a bit steeper than on those other regions, although not dramatically so.  A steeper slope (more energy at larger scales), and a lack of correlation with local sources of kinetic energy (stellar winds and protostellar outflows), suggests that energy is injected on large (100s of pc to kpc) scales.  This agrees with the analysis of
high-J CO and far-infrared line emission at low spatial resolution in 30~Doradus \citet{lee_30dor}, and the existence of kpc-scale colliding filaments in the region \citep{fukui_30dor_formation}.

Clumps and cores when analyzed as entire objects with {\tt clumpfind} lie in a similar part of size-linewidth-mass parameter space as Milky Way and other molecular clouds, and are consistent with free-fall collapse or virial equilibrium with moderate external pressure.  A significant fraction of dendrogram structures and filaments
have mass surface densities $\Sigma_{LTE}$ or line masses $M_l$ in excess of the corresponding quantities if gravitational and kinetic energies were in balance.
The discrepancy can be resolved if the \CO{13}/H$_2$ abundance is a factor of 5 or more {\it higher} than the value that was assumed here.  Alternately, one physical explanation could be that the small scale structures and filaments in this cloud have significant magnetic support against gravity.

The upper end of the ``core'' mass distribution is consistent with a power-law with the same slope as the stellar initial mass function.
Slopes are
fitted directly to the differential mass distribution as well as using
the aforementioned Hill maximum likelihood estimator.  The systematic
biases of direct fitting are more evident than when fitting the
cumulative distribution.  The best estimate slope of -1.3$\pm$0.15 is
consistent with the stellar IMF slope -1.35.
The shape of the mass distribution including the fitted slope does not change if only a subset of cores that have large $\Sigma/\Sigma_{vir}$ are considered; the cores that are more likely to collapse and form stars according to our stability measurement do not have a statistically different mass distribution than those that are less likely to collapse.  This fact, and the fact that the most reliably measured part of the mass distribution has the same slope as the stellar IMF, support the notion that the stellar IMF is predetermined by the characteristics of turbulent fragmentation in the pre-collapse molecular cloud.
A shallower mass function (more massive stars) has been found for the main-sequence massive stars in 30Doradus \citep{science}.  The next generation of star formation in 30Dor-10 will either not have that massive stellar excess, or the stellar mass distribution is not predetermined by the current core mass function.

\acknowledgements

\software{CASA \citep[v4.3.1; v4.5; v5.1.0][]{casa}, Kapteyn \citep{kapteyn}, filfinder \citep{filfinder}, quickclump (\url{https://github.com/vojtech-sidorin/quickclump/}), clumpfind \citep{clumpfind,clumpfind_ascl}, cprops \citep{cprops,cprops_ascl}, dendrograms \citep{dendro}}

This paper makes use of the following ALMA data:
ADS/JAO.ALMA\#2011.0.00471.S, ADS/JAO.ALMA\#2013.1.00346,
ALMA is a partnership of ESO
(representing its member states), NSF (USA) and NINS (Japan),
together with NRC (Canada) and NSC and ASIAA (Taiwan), in
cooperation with the Republic of Chile. The Joint ALMA Observatory
is operated by ESO, AUI/NRAO and NAOJ. The National Radio Astronomy
Observatory is a facility of the National Science Foundation
operated under cooperative agreement by Associated Universities,
Inc.
This research has made use of NASA’s Astrophysics Data System Bibliographic Services.
RI was partially supported during this work by NSF AST-1312902.

% wavelets http://adsabs.harvard.edu/abs/2017arXiv171106663V

\end{document}